\documentclass[floatfix,aps,prd,nofootinbib,notitlepage,superscriptaddress]{revtex4-1}
\pdfoutput=1
\usepackage{graphicx}
\usepackage{amsmath}
\allowdisplaybreaks
\usepackage{amssymb}
\renewcommand\arraystretch{1.5}
\usepackage{slashed}
\usepackage[utf8]{inputenc}
\usepackage[colorlinks=true,linkcolor=blue,bookmarksopen,bookmarksnumbered]{hyperref}
\usepackage{color}
\usepackage[usenames,dvipsnames]{xcolor}
\usepackage{soul}
\usepackage{units}
\usepackage{rotating}
\usepackage{hhline,multirow,tabularx}
\usepackage{bm}
\usepackage{hyperref}
\usepackage[export]{adjustbox}
\usepackage{mdframed}
\usepackage{comment}
\usepackage{natbib}

\newcommand*{\FigPath}{./figures}%

\def\bea#1\eea{\begin{align}#1\end{align}} 

\newcommand{\bef}{\begin{figure}[htb]\centering}
\newcommand{\eef}{\end{figure}}

\newcommand{\nn}{\nonumber}

\def\<{\langle}
\def\>{\rangle}

\def\({\left(}
\def\[{\left[}
\def\){\right)}
\def\]{\right]}

\def\cos{\hbox{cos}}
\def\sin{\hbox{sin}}
\def\ln{\hbox{ln}}
\def\log{\hbox{log}}

\def\Tr{\hbox{Tr}}

\usepackage{lipsum}
\begin{document}
	
\title{The Sivers Asymmetry in Hadronic Dijet Production}
	
\author{Zhong-Bo Kang}
\email{zkang@physics.ucla.edu}
\affiliation{Department of Physics and Astronomy, University of California, Los Angeles, California 90095, USA}
\affiliation{Mani L. Bhaumik Institute for Theoretical Physics, University of California, Los Angeles, California 90095, USA}
\affiliation{Center for Frontiers in Nuclear Science, Stony Brook University, Stony Brook, New York 11794, USA}

\author{Kyle Lee}
\email{kylelee@lbl.gov}
\affiliation{Nuclear Science Division, Lawrence Berkeley National Laboratory, Berkeley, California 94720, USA}
\affiliation{C.N. Yang Institute for Theoretical Physics, Stony Brook University, Stony Brook, New York 11794, USA}
\affiliation{Department of Physics and Astronomy, Stony Brook University, Stony Brook, New York 11794, USA}

\author{Ding Yu Shao}
\email{dingyu.shao@cern.ch}
\affiliation{Department of Physics and Astronomy, University of California, Los Angeles, California 90095, USA}
\affiliation{Mani L. Bhaumik Institute for Theoretical Physics, University of California, Los Angeles, California 90095, USA}
\affiliation{Center for Frontiers in Nuclear Science, Stony Brook University, Stony Brook, New York 11794, USA}

\author{John Terry}
\email{johndterry@physics.ucla.edu}
\affiliation{Department of Physics and Astronomy, University of California, Los Angeles, California 90095, USA}
\affiliation{Mani L. Bhaumik Institute for Theoretical Physics, University of California, Los Angeles, California 90095, USA}


\begin{abstract}
We study the single spin asymmetry in the back-to-back dijet production in transversely polarized proton-proton collisions. Such an asymmetry is generated by the Sivers functions in the incoming polarized proton. We propose a QCD formalism in terms of the transverse momentum dependent parton distribution functions, which allow us to resum the large logarithms that arise in the perturbative calculations. We make predictions for the Sivers asymmetry of hadronic dijet production at the kinematic region that is relevant to the experiment at the Relativistic Heavy Ion Collider~(RHIC). We further compute the spin asymmetries in the selected positive and negative jet charge bins, to separate the contributions from $u$- and $d$-quark Sivers functions. We find that both the sign and size of our numerical results are roughly consistent with the preliminary results from the STAR collaboration at the RHIC. 
\end{abstract}
\maketitle

\section{Introduction}\label{Introduction}
Exploring transverse momentum dependent parton distribution functions (TMD PDFs) has become one of the major research topics in hadron physics in recent years~\cite{Boer:2011fh}. TMD PDFs provide three-dimensional (3D) imaging of the nucleon in both the longitudinal and transverse momentum space, which is one of the scientific pillars at the future Electron-Ion Collider~\cite{Accardi:2012qut}. Such 3D imaging of the nucleon offers novel insights into the highly nontrivial non-perturbative QCD dynamics and correlations~\cite{Aidala:2020mzt}. 

Sivers function is one of the most studied TMD PDFs in the community. It describes the distribution of unpolarized partons inside a transversely polarized nucleon, through a correlation between the transverse spin of the nucleon and the transverse momentum of the parton with respect to the nucleon's moving direction. The Sivers function was first introduced by Sivers in 1990s~\cite{Sivers:1989cc,Sivers:1990fh} to describe the large single transverse spin asymmetries observed in single inclusive particle production in hadronic collisions, see e.g.~\cite{Antille:1980th,Adams:1991rw}. Since then, large single spin asymmetries have also been consistently observed in proton-proton collisions in high energy experiments at the Relativistic Heavy Ion Collider (RHIC)~\cite{Adams:2003fx,Arsene:2008aa,Abelev:2008af,Adamczyk:2012xd,Adare:2013ekj,Adamczyk:2017wld}. On the theoretical side, understanding the precise origin of such large spin asymmetries has triggered extensive research in the QCD community~\cite{Kane:1978nd,Qiu:1991pp,Kouvaris:2006zy,Kang:2010zzb,Kang:2011hk,Metz:2012ct,Gamberg:2013kla,Kanazawa:2014dca,Gamberg:2017gle}. The difficulty in understanding such asymmetries for single hadron production (such as pions) in proton-proton collisions lies in the fact that they could receive contributions from many different correlations. Besides Sivers type correlations, whose collinear version is referred to as the Qiu-Sterman function \cite{Qiu:1991pp,Qiu:1991wg} in the incoming nucleon, there could also be similar correlations in the hadronization process when the parton fragments into the hadrons \cite{Kang:2010zzb,Metz:2012ct,Kanazawa:2010au,Kanazawa:2014dca,Gamberg:2017gle}. See \cite{Cammarota:2020qcw} for a recent development along this direction. 

Simultaneously the Sivers asymmetry has also been studied in semi-inclusive deep inelastic scattering (SIDIS) by HERMES collaboration at DESY~\cite{Airapetian:2009ae,Airapetian:2020zzo}, COMPASS collaboration at CERN~\cite{Adolph:2012sp,Adolph:2016dvl}, and Jefferson Lab~\cite{Qian:2011py}. Because of the semi-inclusive nature of the process, one can isolate the contribution from the Sivers function via different azimuthal angular modulations~\cite{Bacchetta:2006tn}. One of the remarkable and unique properties of the Sivers functions is its non-universality nature. For example, based on parity and time-reversal invariance of QCD, one can show that quark Sivers functions in SIDIS are opposite to those in the Drell-Yan  process~\cite{Brodsky:2002rv,Collins:2002kn,Boer:2003cm}. Such a sign change has been studied and confirmed experimentally~\cite{Aghasyan:2017jop,Adamczyk:2015gyk,Kang:2009bp,Anselmino:2016uie}, though additional work remains to be done to quantify the change in more details~\cite{Aschenauer:2016our}.

Sivers effect has been continuously studied in proton-proton collisions at the RHIC. In order to eliminate the contributions from the spin correlations in the fragmentation process, the Sivers asymmetry for jet production processes has been explored in the experiment~\cite{Bland:2013pkt,Adamczyk:2017wld,Abelev:2007ii}. In particular, back-to-back dijet production in transversely polarized proton-proton collisions was proposed by Boer and Vogelsang in 2003 as a unique opportunity at the RHIC~\cite{Boer:2003tx}. Active investigation has been performed both experimentally~\cite{Abelev:2007ii} and theoretically~\cite{Bomhof:2007su,Vogelsang:2007jk,Qiu:2007ar}. On the experimental side, the Sivers asymmetry for dijet production was found to be quite small, largely due to the cancellation between $u$- and $d$-quark Sivers functions, which have similar size but opposite sign~\cite{Echevarria:2014xaa,Bacchetta:2020gko,Cammarota:2020qcw}. On the theoretical side, dijet production in proton-proton collisions is also subject to TMD factorization breaking~\cite{Collins:2007nk,Rogers:2010dm}. These have slowed down the efforts in the detailed study of the Sivers effect in the dijet production. 

Recently, there have been renewed experimental and theoretical interests for jet production processes. Experimentally, the STAR collaboration at the RHIC is analyzing the new data for dijet Sivers asymmetry, and is exploring a novel method based on a charge weighting method in separating the contributions from individual $u$ and $d$-quark Sivers functions~\cite{bnltalk}. The PHENIX collaboration at the RHIC is exploring the TMD factorization breaking effects via back-to-back dihadron and photon-hadron production in proton-proton collisions~\cite{Adare:2016bug,Aidala:2018bjf}. Theoretically, there have been efforts in performing QCD resummation in back-to-back dijet~\cite{Sun:2014gfa,Sun:2015doa} and vector boson-jet production~\cite{Buffing:2018ggv,Chien:2019gyf,Chien:2020hzh}. At the same time, a theoretical framework has been developed to study spin asymmetries in specific jet charge bin~\cite{Kang:2020fka}, which would facilitate the analysis of the dijet spin asymmetries by the STAR collaboration. In light of all these activities, we set out to develop a resummation formalism for studying the Sivers asymmetry in back-to-back dijet production in transversely polarized proton-proton collisions. We make predictions for the dijet Sivers asymmetry in the kinematics relevant to the RHIC energy, to be compared with the experimental measurement in the near future. 

The rest of the paper is organized as follows. In Section~\ref{Factorization}, we summarized the QCD formalism for dijet production in both unpolarized and polarized scatterings, and we provide a few remarks about our formalism. In Section~\ref{Hard Functions}, we provide a procedure and demonstrate how to compute the process-dependent polarized hard functions in the color matrix form. In Section~\ref{sec:resummation}, we present the renormalization group evolution of all the relevant functions in our formalism, and we provide the final resummation formula. Section~\ref{Phenomenology} is devoted to the phenomenological studies, where we make predictions for dijet Sivers asymmetry in the kinematic region relevant to the experiment at the RHIC. Since we are mainly interested in the Sivers asymmetry in the forward rapidity region where quark contributions dominate, we consider only the quark Sivers contribution and neglect the gluon Sivers contribution. We summarize our paper in Section~\ref{Conclusions}.

\section{QCD formalism for dijet production}\label{Factorization}
In this paper, we study back-to-back dijet production in transversely polarized proton-proton collisions in the center-of mass frame, 
\bea
p(P_{A}, \vec{S}_\perp) + p(P_B) \to J_1(y_c, \vec{P}_{1\perp}) + J_2(y_d, \vec{P}_{2\perp})+X\,,
\eea
where the polarized proton with the momentum $P_A$ and the transverse spin $\vec{S}_\perp$ is moving in the $+z$-direction, while the unpolarized proton with the momentum $P_B$ is moving in the $-z$-direction, and we have the center-of-mass energy $s=(P_A+P_B)^2$. The produced two jets $J_1$ and $J_2$ have rapidities $y_{c,d}$ and transverse momenta $\vec{P}_{1\perp}$ and $\vec{P}_{2\perp}$, respectively. 
\begin{figure}[h!tb]
\begin{center}
    \includegraphics[width=0.35\linewidth]{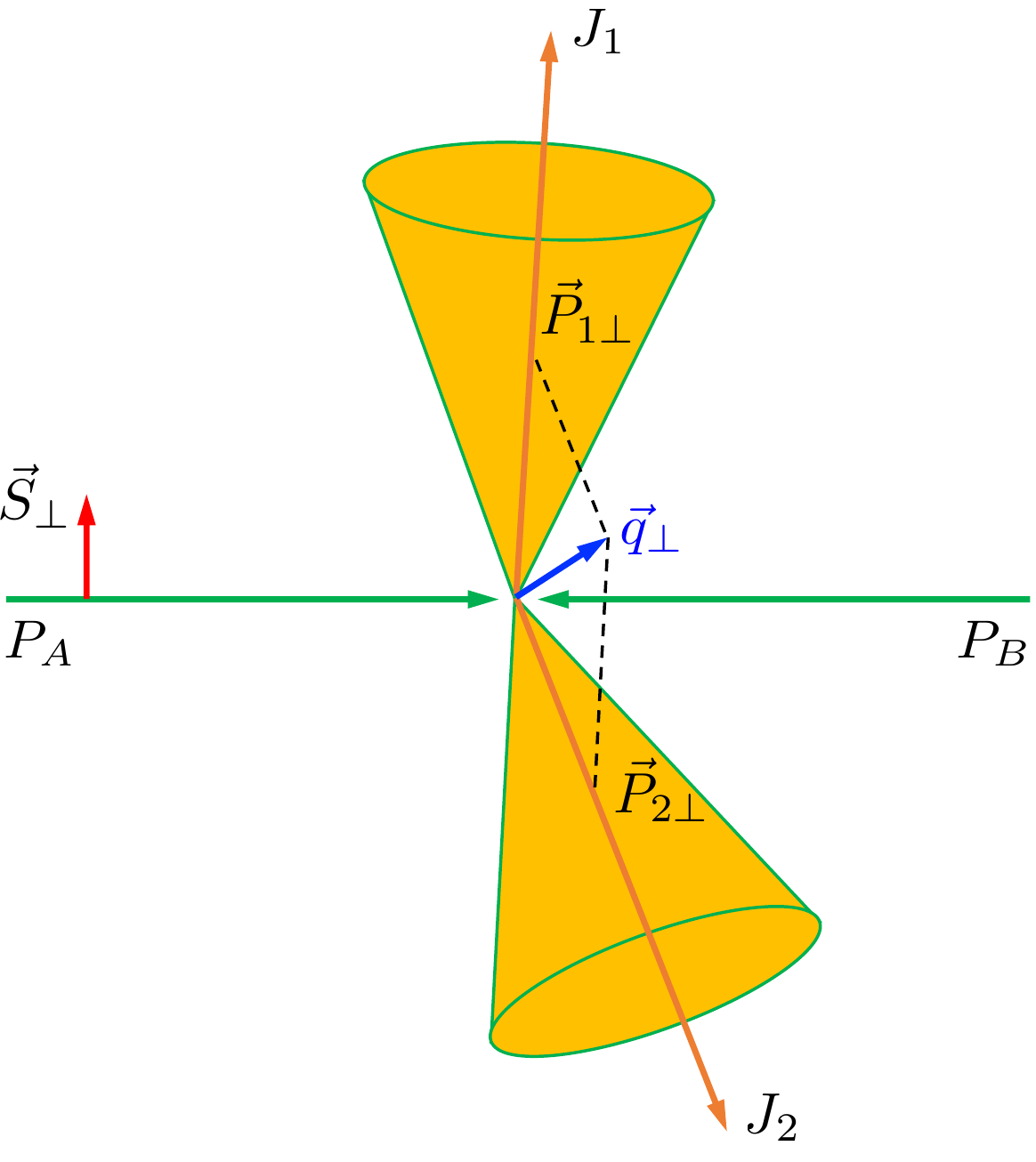}
\end{center}
\caption{Illustration of back-to-back dijet production in transversely polarized proton-proton collisions: $p(P_{A}, \vec{S}_\perp) + p(P_B) \to J_1(y_c, \vec{P}_{1\perp}) + J_2(y_d, \vec{P}_{2\perp})+X$. The polarized proton with momentum $P_A$ and transverse spin $\vec{S}_\perp$ is moving in $+z$-direction, while the unpolarized proton with momentum $P_B$ is moving in $-z$-direction. We have jet rapidities $y_{c,d}$ and transverse momenta $\vec{P}_{1\perp}$ and $\vec{P}_{2\perp}$, respectively. The dijet transverse momentum imbalance is defined as $\vec{q}_\perp = \vec{P}_{1\perp}+\vec{P}_{2\perp}$. Sivers asymmetry is generated due to the correlation between $\vec{S}_\perp$ and $\vec{q}_\perp$.}
\label{fig:dijet_fig}
\end{figure}
These jets will be reconstructed via a suitable jet algorithm~\cite{Cacciari:2011ma} and in the rest of the paper, we consider both of them to be anti-$k_T$ jets with jet radii $R$. In order to access the transverse motion of the partons inside the protons, we concentrate in the back-to-back region where the transverse momentum imbalance $q_\perp$ is small. Here we define the average transverse momentum $P_\perp$ of the two jets and the transverse momentum imbalance $\vec{q}_\perp$ as follows 
\bea
P_\perp = |\vec{P}_{1\perp} - \vec{P}_{2\perp}|/2\,,
\qquad
\vec{q}_\perp = \vec{P}_{1\perp} + \vec{P}_{2\perp}\,,
\eea
where one has $q_\perp \ll P_\perp$ in the back-to-back region. The production of such back-to-back dijets is illustrated in Fig.~\ref{fig:dijet_fig}. In the transversely polarized proton-proton collisions, the transverse spin vector $\vec{S}_\perp$ of the incoming proton and the transverse momentum imbalance $\vec{q}_\perp$ of the two jets will be correlated, as advocated in~\cite{Boer:2003tx}. This correlation is accounted for in the Sivers function, which leads to a $\sin(\phi_q - \phi_S)$-azimuthal modulation in the cross section between $\phi_q$ and $\phi_S$, the azimuthal angles of $\vec{q}_\perp$ and $\vec{S}_\perp$, respectively.
Below we summarize the factorized formalisms for dijet production in both unpolarized and polarized proton-proton collisions, and we provide more details for the relevant ingredients in the next section. 

\subsection{Dijet unpolarized cross section}
In the back-to-back region where $q_\perp\ll P_\perp$, within the framework of soft-collinear effective theory (SCET)~\cite{Bauer:2000ew,Bauer:2000yr,Bauer:2001ct,Bauer:2001yt,Bauer:2002nz}, one can write down a factorized form for the unpolarized differential cross section
\begin{align}
\label{eq:momunpol}
\frac{d\sigma}{dy_c dy_d d P_\perp^2 d^2 \vec q_\perp} =& \sum_{abcd} \frac{1}{16\pi^2 \hat{s}^2} \frac{1}{N_{\rm init}} \frac{1}{1+\delta_{cd}} \int_\perp \,x_a f^{\rm unsub}_a(x_a,k_{a\perp},\mu,\nu)\,x_b f^{\rm unsub}_b(x_b,k_{a\perp},\mu,\nu)  \notag \\
&\times {\rm Tr}\left[ \bm{S}_{ab\rightarrow cd}(\lambda_{\perp},\mu,\nu)\cdot
\bm{H}_{ab\rightarrow cd}(P_{\perp},\mu) \right] J_c(P_\perp R,\mu)  S^{\rm cs}_c(k_{c\perp}, R,\mu) J_d(P_\perp R,\mu)  S^{\rm cs}_d(k_{d\perp}, R,\mu)\,,
\end{align}
where $\hat s = x_a x_b s$ is the partonic center-of-mass energy, $N_{\rm init}$ is the corresponding spin- and color-averaged factor for each channel, while $1/(1+\delta_{cd})$ arises from the symmetry factor due to identical partons in the final state. We have used the following short-hand notation
\begin{align}
    \int_{\perp} = \int d^2\vec{k}_{a\perp} d^2\vec{k}_{b\perp} d^2\vec{k}_{c\perp} d^2\vec{k}_{d\perp}  d^2\vec{\lambda}_{\perp} \delta^{(2)}(\vec{k}_{a\perp}+\vec{k}_{b\perp}+\vec{k}_{c\perp}+\vec{k}_{d\perp}+\vec{\lambda}_{\perp}-\vec{q}_{\perp})\,.
\end{align}
In Eq.~\eqref{eq:momunpol}, $f^{\rm unsub}_a(x_a, k_{a\perp},\mu,\nu)$ and $f^{\rm unsub}_b(x_b, k_{b\perp},\mu,\nu)$ are the so-called unsubtracted TMD PDFs, which carry the longitudinal momentum fractions $x_{a,b}$ and the transverse momenta $k_{a\perp}$ and $k_{b\perp}$ with respect to their corresponding proton. In our process, we have
\bea
x_a = \frac{P_\perp}{\sqrt{s}}\left(e^{y_c} + e^{y_d}\right)\,,
\qquad
x_b = \frac{P_\perp}{\sqrt{s}}\left(e^{-y_c} + e^{-y_d}\right)\,,
\eea
where $y_c$, $y_d$ are the rapidities of the two leading jets.

After performing Fourier transform for Eq.~\eqref{eq:momunpol}, we obtain the factorized formula in the coordinate $b$-space as follows
\begin{align}
\label{eq:bunpol}
\frac{d\sigma}{dy_c dy_d d P_\perp^2 d^2\vec q_\perp } =& \sum_{abcd} \frac{1}{16\pi^2 \hat{s}^2} \frac{1}{N_{\rm init}} \frac{1}{1+\delta_{cd}} \int \frac{d^2\vec{b}}{(2\pi)^2}\,e^{i\vec{q}_\perp\cdot\vec{b}} \,x_a f^{\rm unsub}_a(x_a,b,\mu,\nu)\, x_b f^{\rm unsub}_b(x_b,b,\mu,\nu) \notag \\
&\times {\rm Tr}\left[ \bm{S}_{ab\rightarrow cd}(b,\mu,\nu)\cdot
\bm{H}_{ab\rightarrow cd}(P_{\perp}, \mu) \right]   J_c(P_\perp R,\mu)  S^{\rm cs}_c(b,R,\mu) J_d(P_\perp R,\mu)  S^{\rm cs}_d(b,R,\mu)\,,
\end{align}
where $f^{\rm unsub}_a(x_a,b,\mu,\nu)$ and $f^{\rm unsub}_b(x_b,b,\mu,\nu)$ are the Fourier transform of  $f^{\rm unsub}_a(x_a, k_{a\perp},\mu,\nu)$ and $f^{\rm unsub}_b(x_b, k_{b\perp},\mu,\nu)$, respectively. On the other hand, $\bm{H}_{ab\rightarrow cd}(P_\perp, \mu)$ is the hard function, while $\bm{S}_{ab\to cd}(b, \mu, \nu)$ is a global soft function. Note that both the hard function $\bm{H}_{ab\rightarrow cd}$ and the global soft function $\bm{S}_{ab\to cd}$ are expressed in the matrix form in the color space and the trace ${\rm Tr}[\cdots]$ is over the color. Such factorization of the hard and soft function into matrix form is essential to capture evolution effects between the hard scale $\sim P_\perp$ and the imbalance scale $\sim q_\perp$ \cite{Kidonakis:1998nf}. Here $\mu$ and $\nu$ denotes renormalization and rapidity scales, separately. The rapidity scale $\nu$ arises because both the TMD PDFs and the global soft functions have rapidity divergence~\cite{Chiu:2011qc,Chiu:2012ir}, which are canceled between them as demonstrated below. This cancellation allows us to define rapidity divergence independent $\tilde{\bm{S}}_{ab\to cd}(b,\mu)$ by 
\bea
\bm{S}_{ab\to cd}(b,\mu,\nu) = \tilde{\bm{S}}_{ab\to cd}(b,\mu)S_{ab} (b,\mu,\nu)\,,
\label{eq:soft-fac}
\eea
where $S_{ab} (b,\mu,\nu)$ is the standard soft function appearing in usual Drell-Yan and SIDIS processes. This explicit redefinition allows us to subtract the rapidity divergence from the unsubtracted TMD PDFs to define the standard TMD PDFs $f_i(x_i,b,\mu)$ that are free of rapidity divergence as~\cite{Collins:2011zzd}
\bea
f^{\rm unsub}_a(x_a,b,\mu,\nu)\, f^{\rm unsub}_b(x_b,b,\mu,\nu) \,S_{ab} (b,\mu,\nu) = f_a(x_a,b,\mu)\, f_b(x_b,b,\mu)\,.
\label{eq:proper}
\eea
Note that the properly-defined TMD PDFs $f_a(x_a,b,\mu)$ and  $f_b(x_b,b,\mu)$ are no longer subject to the rapidity divergence and this is why there are no explicit $\nu$-dependence in the arguments any more. Such properly-defined unpolarized TMD PDFs are the same as those probed in the standard SIDIS and Drell-Yan processes. 

The jet functions $J_c(P_\perp R,\mu)$ and $J_d(P_\perp R,\mu)$ in Eq.~\eqref{eq:bunpol} describe the creation of anti-$k_T$ jets from the partons $c$ and $d$, respectively. Finally, $S^{\rm cs}_c(k_{c\perp}, R,\mu)$ and $S^{\rm cs}_d(k_{d\perp}, R,\mu)$ are the collinear-soft functions. They describe soft gluon radiation with separations of order $R$ along the jet direction, which can resolve the substructure of the jet. 
If one performs the integration over the azimuthal angle of the vector $\vec{b}$, we obtain the following expression
\bea
\frac{d\sigma}{dy_c dy_d d P_\perp^2 d^2\vec q_\perp } =& \sum_{abcd} \frac{1}{16\pi^2 \hat{s}^2} \frac{1}{N_{\rm init}} \frac{1}{1+\delta_{cd}} \frac{1}{2\pi}\int_0^{\infty} db \,b\,J_0(q_\perp b) \,x_a f_a(x_a,b,\mu)\, x_b f_b(x_b,b,\mu) \notag \\
&\times {\rm Tr}\left[ \tilde{\bm{S}}_{ab\rightarrow cd}(b,\mu)\cdot
\bm{H}_{ab\rightarrow cd}(P_{\perp}, \mu) \right]   J_c(P_\perp R,\mu)  S^{\rm cs}_c(b,R,\mu) J_d(P_\perp R,\mu)  S^{\rm cs}_d(b,R,\mu)\,,
\label{eq:unpol-final}
\eea
where $J_0$ is the Bessel function of order zero. 

\subsection{Dijet Sivers asymmetry}
In the transversely polarized proton-proton collisions, the Sivers function will lead to a spin asymmetry in the cross section when one flips the transverse spin of the incoming proton. We thus define the difference in the cross section as $d\Delta\sigma(S_\perp) = \left[d\sigma(S_\perp) - d\sigma(-S_\perp)\right]/2$. One can write down a similar factorized formula for such a spin-dependent differential cross section following Eq.~\eqref{eq:momunpol}, and it is given by
\bea
\label{eq:spin-cross}
\frac{d\Delta \sigma(S_\perp)}{dy_c dy_d d P_\perp^2 d^2 \vec q_\perp} =&  \sum_{abcd} \frac{1}{16\pi^2 \hat{s}^2} \frac{1}{N_{\rm init}} \frac{1}{1+\delta_{cd}} \int_\perp \,
\frac{1}{M} \epsilon_{\alpha\beta} \,S_\perp^\alpha \,k_{a\perp}^\beta \,x_a f_{1T}^{\perp a,\, {\rm unsub}}(x_a, k_{a\perp}, \mu, \nu)
\,x_b f^{\rm unsub}_b(x_b,k_{a\perp},\mu, \nu)  
\notag \\
&\times {\rm Tr}\left[ \bm{S}_{ab\rightarrow cd}(\lambda_{\perp},\mu, \nu)\cdot
\bm{H}_{ab\rightarrow cd}^{\rm Sivers}(P_{\perp},\mu) \right] J_c(P_\perp R,\mu)  S^{\rm cs}_c(k_{c\perp}, R,\mu) J_d(P_\perp R,\mu)  S^{\rm cs}_d(k_{d\perp}, R,\mu)\,,
\eea
where $\epsilon_{\alpha\beta}$ is a two-dimensional asymmetric tensor with $\epsilon_{12}=+1$, and we have replaced the unpolarized TMD PDF in Eq.~\eqref{eq:momunpol} by the Sivers function in the above equation following the so-called Trento convention~\cite{Bacchetta:2004jz},
\bea
f_a^{\rm unsub}(x_a, k_{a\perp},\mu, \nu) \rightarrow \frac{1}{M} \epsilon_{\alpha\beta} \,S_\perp^\alpha \,k_{a\perp}^\beta \, f_{1T}^{\perp a,\,{\rm unsub}}(x_a, k_{a\perp}, \mu, \nu)\,.
\label{eq:siver-sidis}
\eea
Note that we have also assumed that the global soft function $\bm{S}_{ab\rightarrow cd}(\lambda_{\perp},\mu, \nu)$ stays the same as that of the unpolarized collisions in Eq.~\eqref{eq:momunpol}. Although this is a reasonable assumption since the soft gluon radiation should be spin-independent~\cite{Kelley:2010fn,Echevarria:2015usa}, this has to be carefully checked. In fact, Ref.~\cite{Liu:2020jjv} shows in explicit calculations at one-loop level that soft functions in the polarized case can be different from the unpolarized counterpart beyond leading logarithmic accuracy, which is an indication of TMD factorization breaking. In this respect, our starting point Eq.~\eqref{eq:spin-cross} will be the best assumption at hand that takes a factorized form. We show the RG consistency for this factorized form, and we also demonstrate how we derive the process-dependent hard functions $\bm{H}_{ab\rightarrow cd}^{\rm Sivers}(P_{\perp},\mu)$ for the polarized scattering. We leave a detailed study on the numerical impact of any TMD factorization breaking effects for future investigation.

Performing Fourier transform from the transverse momentum space into the $b$-space, we obtain
\bea
\frac{d\Delta\sigma(S_\perp)}{dy_c dy_d d P_\perp^2 d^2\vec q_\perp } =& \sum_{abcd} \frac{1}{16\pi^2 \hat{s}^2} \frac{1}{N_{\rm init}} \frac{1}{1+\delta_{cd}} \epsilon_{\alpha\beta} \,S_\perp^\alpha \int \frac{d^2\vec{b}}{(2\pi)^2}\,e^{i\vec{q}_\perp\cdot\vec{b}} \,x_a f_{1T}^{\perp \, a(\beta)}(x_a, b,\mu)\, x_b f_b(x_b,b,\mu) \notag \\
&\times {\rm Tr}\left[ \tilde{\bm{S}}_{ab\rightarrow cd}(b,\mu)\cdot
\bm{H}_{ab\rightarrow cd}^{\rm Sivers}(P_{\perp}, \mu) \right]   J_c(P_\perp R,\mu)  S^{\rm cs}_c(b,R,\mu) J_d(P_\perp R,\mu)  S^{\rm cs}_d(b,R,\mu)\,,
\label{eq:b-pol}
\eea
where we have already used Eq.~\eqref{eq:soft-fac} to rewrite the unsubtracted unpolarized TMD PDF and Sivers function in terms of the properly defined versions which are free of rapidity divergence. Here $f_{1T}^{\perp \, a(\beta)}(x_a, b,\mu)$ is the Fourier transform of the Sivers function,
\bea
f_{1T}^{\perp \, a(\beta)}(x_a, b,\mu) =\,& \frac{1}{M} \int d^2\vec{k}_{a\perp} \,e^{-i\vec{k}_{a\perp}\cdot \vec{b}} \,k_{a\perp}^{\beta}f_{1T}^{\perp \, a }(x_a, k_{a\perp},\mu)\,,
\notag\\
\equiv\,&\left(\frac{ib^\beta}{2}\right)
\hat{f}_{1T}^{\perp \, a}(x_a, b,\mu) \,,
\label{eq:siv-b-def}
\eea
where we have used the fact that the integration in the first line would be proportional to $b^\beta$, and we thus factored $b^\beta$ out explicitly in the second line~\footnote{To make the matching coefficient normalized to 1 at the lowest order in Eq.~\eqref{eq:siv-collinear}, we include the additional factor of $i/2$ in Eq.~\eqref{eq:siv-b-def}.}. The remaining part of the Sivers function is now denoted as $\hat{f}_{1T}^{\perp \, a}(x_a, b,\mu)$. Note that for the same reason as explained below Eq.~\eqref{eq:proper}, we do not have the rapidity $\nu$-dependence in the above equation. It is also instructive to emphasize that $\hat{f}_{1T}^{\perp \, a}(x_a, b, \mu)$ follows the same TMD evolution equations as the unpolarized TMD PDF $f_a(x_a, b, \mu)$, which enables us to evolve the Sivers function from some initial scale $\mu_0$ to the relevant scale $\mu$. On the other hand, at the initial scale $\mu_0$, the unpolarized TMD PDF $f_a(x_a, b, \mu_0)$ can be expanded in terms of the collinear PDFs $f_a(x_a, \mu_0)$. At a specific scale $\mu_b = b_0/b$ with $b_0=2e^{-\gamma_E}$, we have
\bea
f_a(x_a, b, \mu_b) = \int_{x_a}^1\frac{dx}{x} \,C_{a\leftarrow i}\left(\frac{x_a}{x}, \mu_b\right) \, f_{i}(x, \mu_b)\,,
\eea
where the coefficient $C_{a\leftarrow i}$ can be found in e.g. Refs.~\cite{Collins:2011zzd,Aybat:2011zv}. Likewise, Sivers function $\hat{f}_{1T}^{\perp \, a}(x_a, b, \mu)$ can be further matched onto the collinear twist-three Qiu-Sterman function $T_{a,F}(x_1, x_2, \mu)$. At the scale $\mu_b$, one has the following expression for quark Sivers functions
\bea
\hat{f}_{1T}^{\perp \, q}(x_a, b,\mu_b) = \int_{x_a}^1 \frac{dx}{x} \,C^T_{q\leftarrow q'}\left(\frac{x_a}{x}, \mu_b\right) \, T_{q', F}(x, x, \mu_b)\,,
\label{eq:siv-collinear}
\eea
where the matching coefficients at the NLO are given by~\cite{Kang:2011mr,Sun:2013hua,Dai:2014ala,Scimemi:2019gge,Moos:2020wvd}
\begin{align}\label{eq:siver-matching}
C^T_{q\leftarrow q'}\left(x, \mu_b\right) = \delta_{qq'}\left[\delta(1-x) + \frac{\alpha_s(\mu_b)}{2\pi}\left(-\frac{1}{2N_c}\right)(1-x) \right]\,.
\end{align}
We now plug Eq.~\eqref{eq:siv-b-def} into Eq.~\eqref{eq:b-pol}, and integrate over the azimuthal angle of the vector $\vec{b}$, we obtain
\bea
\frac{d\Delta\sigma(S_\perp)}{dy_c dy_d d P_\perp^2 d^2\vec q_\perp } =\,& \sin(\phi_q - \phi_S)\sum_{abcd} \frac{1}{16\pi^2 \hat{s}^2} \frac{1}{N_{\rm init}} \frac{1}{1+\delta_{cd}} \left(-\frac{1}{4\pi}\right) \int_0^\infty db \, b^2\, J_1(q_\perp b)\, x_a \hat{f}_{1T}^{\perp \, a}(x_a, b,\mu)\, x_b f_b(x_b,b,\mu) \notag \\
&\,\times {\rm Tr}\left[ \tilde{\bm{S}}_{ab\rightarrow cd}(b,\mu)\cdot
\bm{H}_{ab\rightarrow cd}^{\rm Sivers}(P_{\perp}, \mu) \right]   J_c(P_\perp R,\mu)  S^{\rm cs}_c(b,R,\mu) J_d(P_\perp R,\mu)  S^{\rm cs}_d(b,R,\mu)\,,
\label{eq:pol-final}
\eea
where $J_1$ is the Bessel function of order one, and we have used the identity
\bea
\epsilon_{\alpha\beta} S_\perp^\alpha\, \hat{q}_\perp^\beta = \sin(\phi_q - \phi_S)\,,
\eea
with $\hat q_\perp$ the unit vector along the direction of the imbalance $\vec{q}_\perp$. In general, the so-called single spin asymmetry (the Sivers asymmetry) $A_N$ for dijet production will be then given by 
\bea
A_N = \left.\frac{d\Delta\sigma(S_\perp)}{dy_c dy_d d P_\perp^2 d^2\vec q_\perp }\right/\frac{d\sigma}{dy_c dy_d d P_\perp^2 d^2\vec q_\perp }\,.
\eea

Finally, since the Sivers function is not universal, one has to carefully include those non-universality or process-dependence into the above formalism~\cite{Bacchetta:2005rm,Bomhof:2006dp,Bomhof:2007su,Qiu:2007ey,Vogelsang:2007jk,Qiu:2007ar,Collins:2011zzd}. We have chosen to include all such process-dependence into the hard function $\bm{H}_{ab\rightarrow cd}^{\rm Sivers}(P_{\perp},\mu)$, and this way the Sivers functions in Eq.~\eqref{eq:pol-final} are the same as those probed in the SIDIS process. We explain in details how we derive the hard functions $\bm{H}_{ab\rightarrow cd}^{\rm Sivers}$ for different partonic processes in the next section. 

\subsection{Remarks}\label{sec:remark}
We will provide detailed expressions and discuss the evolution of all the relevant functions in the next section. Here, let us emphasize the following points on our factorized formalism:

\begin{itemize}
  \item Eqs.~\eqref{eq:bunpol} and \eqref{eq:b-pol} are our proposed factorized formulas for dijet production in unpolarized and transversely polarized proton-proton collisions, respectively. They are the essential theoretical formalism we are using in the phenomenology section to compute the dijet Sivers asymmetry, which can be compared with the experimental data at the RHIC. 
  
  \item It is important to emphasize that we have derived both  Eqs.~\eqref{eq:bunpol} and \eqref{eq:b-pol} within the SCET framework, in which the Glauber mode is absent. However, it is well-known that the inclusion of the Glauber modes will lead to factorization breaking. The factorization violation effects from Glauber gluon exchanging diagrams between two incoming nucleons have been discussed in~\cite{Collins:2007nk,Rogers:2010dm,Catani:2011st,Forshaw:2012bi}. In principle, such effects can be systematically accounted for in SCET by including explicitly the Glauber mode~\cite{Rothstein:2016bsq}. How exactly this works for dijet production remains to be investigated. In any case, the formalism we presented here would be a good starting point. This formalism incorporates the process dependence of the Sivers functions as outlined in~\cite{Bacchetta:2005rm,Bomhof:2007su,Qiu:2007ey,Vogelsang:2007jk,Qiu:2007ar}, and also properly takes care of the QCD resummation and evolution effects. Thus in this formalism, we are able to study the energy and scale dependence of the Sivers asymmetry as measured in the experiment. 
  
  \item There will be non-global structures from quantum correlations between in-jet and out-of-jet radiations: exclusive jet production will be sensitive on the correlation effects between in-jet and out-of-jet radiations, which is first discovered in~\cite{Dasgupta:2001sh}. The corresponding factorization and resummation formula involves multi-Wilson-line structures~\cite{Becher:2015hka,Becher:2016mmh}, which will give the non-linear evolution equation \cite{Sterman:2004en} for non-global logarithms (NGLs) resummation. The TMD factorization formula including such effects have been given in~\cite{Becher:2017nof,Chien:2019gyf,Kang:2020yqw}. Numerically, the leading-logarithmic NGLs resummation can be solved using parton shower methods~\cite{Dasgupta:2001sh,Dasgupta:2002bw,Balsiger:2019tne,Neill:2018yet} or BMS equations~\cite{Banfi:2002hw,Hatta:2017fwr}. In our phenomenology, we have included the contributions from leading-logarithmic NGLs as discussed in Section~\ref{Phenomenology}.
  
  \item Our formalism for unpolarized dijet production in Eqs.~\eqref{eq:bunpol} is similar to those in~\cite{Sun:2014gfa,Sun:2015doa}. Here, by taking the small-$R$ limit, we refactorize the TMD $R$-dependence soft function~\cite{Sun:2014gfa,Sun:2015doa} as the product of the $R$-independent global TMD soft function and the $R$-dependent collinear-soft function~\cite{Buffing:2018ggv,Chien:2019gyf}. In addition, the $R$-dependent hard function in~\cite{Sun:2014gfa,Sun:2015doa} has been further factorized into a $R$-independent hard function as above and the jet functions which naturally capture all the $R$-dependence. In this regard, the factorized formula presented here is more transparent and intuitive. Such refactorizations are essential to resum logarithms of $R$ for small radius jets. 
  
  \item After performing the refactorization mentioned in the above item, both the single logarithmic anomalous dimensions of the global and collinear-soft function not only depend on the magnitude $|\vec b|$ but also the azimuthal angle $\phi_b$ of the vector $\vec b$ \cite{Buffing:2018ggv,Chien:2019gyf}. Especially, after taking into account QCD evolution effects the $\phi_b$ integral is divergent in some phase space region. In order to regularize such divergences, we can first take $\phi_b$ averaging in both the global and collinear-soft function, and then explicit $\phi_b$ dependence will vanish. Therefore, one can avoid such divergence in the resummation formula directly. This $\phi_b$ averaging method will not change the RG consistency at the one-loop order. The other methods to avoid such divergence have been discussed in  \cite{Chien:2019gyf}, and no significant numerical differences are found at the NLL accuracy. The similar $\phi_b$ averaging methods have also been used in \cite{Zhu:2012ts,Li:2013mia,Angeles-Martinez:2018mqh} to simplify the calculation of the TMD soft function.
\end{itemize}


\section{Hard Functions in unpolarized and polarized scattering}\label{Hard Functions}
In this section, we derive the hard functions for both unpolarized and polarized scatterings, i.e. $\bm{H}_{ab\rightarrow cd}(P_{\perp}, \mu)$ and $\bm{H}_{ab\rightarrow cd}^{\rm Sivers}(P_{\perp}, \mu)$ in Eqs.~\eqref{eq:unpol-final} and \eqref{eq:pol-final}, respectively. They are matrices in the color space. We first review the results for the hard functions $\bm{H}_{ab\rightarrow cd}$ in the unpolarized scattering, which are well-known in the literature, see e.g. Refs.~\cite{Kelley:2010fn,Liu:2014oog}. We then derive the hard function matrices $\bm{H}_{ab\rightarrow cd}^{\rm Sivers}$ in the polarized scattering case. These hard functions properly take into account the process-dependence of the Sivers functions~\cite{Bacchetta:2005rm,Bomhof:2006dp,Bomhof:2007su,Qiu:2007ey,Vogelsang:2007jk,Qiu:2007ar,Collins:2011zzd}. To get started, we define the Mandelstam variables for the partonic scattering process, $a(p_1) + b(p_2) \to c(p_3) + d(p_4)$, as follows
\begin{subequations}
\bea
\hat{s} &= (p_1+p_2)^2 = (p_3+p_4)^2 = 4P_\perp^2 \cosh^2\left(\frac{\Delta y}{2}\right) = x_a x_b s\,,\\
\hat{t} &= (p_1-p_3)^2 = (p_2 - p_4)^2= -2P_\perp^2 e^{-\Delta y/2}\cosh\left(\frac{\Delta y}{2}\right)\,,\\
\hat{u} &= (p_1-p_4)^2 = (p_2 - p_3)^2 = -2P_\perp^2 e^{\Delta y/2}\cosh\left(\frac{\Delta y}{2}\right)\,,
\eea
\end{subequations}
where $\Delta y = y_c-y_d$ is the rapidity difference of the two jets. In the following, the expressions for the hard functions will be written in terms of these Mandelstam variables.

\subsection{Unpolarized Hard Matrices}\label{Upol}
\subsubsection{Four quark subprocesses}
\begin{figure}
  \centering
  \includegraphics[valign = b, height = 1.15in]{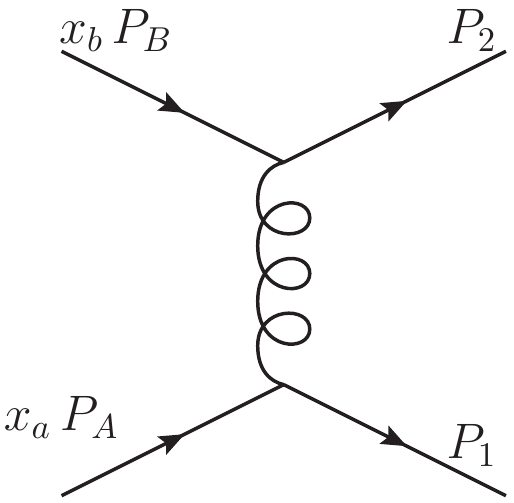}
  \qquad
  \includegraphics[valign = b, height = 1.125in]{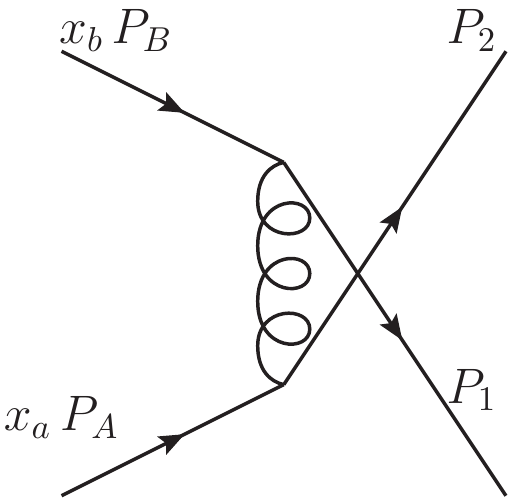}
  \caption{Unpolarized scattering amplitudes for the $qq\rightarrow qq$ subprocess. From left the right, the scattering amplitude is provided for the $t$- and $u$-channel processes.}
  \label{fig:M_upol}
\end{figure}
We start with the partonic subprocesses that involve four quarks, such as $qq\to qq$. In Tab.~\ref{tab:quark-channels}, we organize each of the four quark subprocesses into a color basis. The color basis operators acting on particles $i$ and $j$ are denoted as $\Gamma_{n,ij}$ which are used to generate the hard and soft matrices. For the four quark interactions, two operators, $n = 1, 2$, are required to span the color space. As seen in the table, this results in 12 total color matrices. Using the fact that hard function for the unpolarized case is invariant under the charge conjugation, the bottom row can easily be computed from the top row. Furthermore, once the hard matrices have been calculated for the first column, crossing symmetry can be applied in order to obtain the hard color matrices for the second and third column. It is then only necessary to explicitly calculate the hard matrices for the subprocesses associated with the color basis $\Gamma_{n,31}\Gamma_{n,42}$.
\begin{table}[t!]
\def\arraystretch{2.}
\setlength{\tabcolsep}{6pt}
 \begin{tabular}{||c | c || c | c || c | c ||} 
 \hline
 12 $\rightarrow$ 34 & Color Basis & 12 $\rightarrow$ 34 & Color Basis & 12 $\rightarrow$ 34 & Color Basis \\[0.5ex] 
 \hline
 \hline
 $qq' \rightarrow qq'$ &  & $q\bar{q} \rightarrow q'\bar{q}'$ &  & $q\bar{q}' \rightarrow \bar{q}'q $ &   \\
 $qq' \rightarrow q'q$ & $\Gamma_{n,31}\Gamma_{n,42}$ & $q\bar{q}'\rightarrow q\bar{q}' $ & $\Gamma_{n,21}\Gamma_{n,34}$ & $q\bar{q}  \rightarrow \bar{q}'q'$ & $\Gamma_{n,41}\Gamma_{n,23}$ \\
 $qq' \rightarrow qq $ &                                     & $q\bar{q} \rightarrow q\bar{q}  $ &                                     & $q\bar{q} \rightarrow \bar{q}q   $ &  \\
 \hline
  $\bar{q}\bar{q}' \rightarrow \bar{q}\bar{q}'$ &   & $\bar{q}q \rightarrow \bar{q}'q'$ &   & $\bar{q}q' \rightarrow q'\bar{q} $ &   \\
 $\bar{q}\bar{q}' \rightarrow \bar{q}'\bar{q}$ & $\Gamma_{n,13}\Gamma_{n,24}$ & $\bar{q}q'\rightarrow \bar{q}q' $ & $\Gamma_{n,12}\Gamma_{n,43}$ & $\bar{q}q  \rightarrow q'\bar{q}'$ & $\Gamma_{n,14}\Gamma_{n,32}$ \\
 $\bar{q}\bar{q}' \rightarrow \bar{q}\bar{q} $ &                                     & $\bar{q}q \rightarrow \bar{q}q  $ &                                     & $\bar{q}q \rightarrow q\bar{q}   $ &  \\
 \hline
\end{tabular}
    \caption{The choice of basis for each of the four quark subprocesses. $\Gamma_{n,ij}$ are operators in color space which join the fermion lines $i$ and $j$. For the four quark subprocesses, two operators, $\Gamma_{1,ij}$ and $\Gamma_{2,ij}$, are required to span the color space.}
    \label{tab:quark-channels}
\end{table}
For our calculation, we follow the conventions used in Refs.~\cite{Liu:2014oog,Kelley:2010fn} to choose $\Gamma_{1,ij} = (t^a)_{ij}$ and $\Gamma_{2,ij} = \delta_{ij}$, so that the color basis is spanned by the orthogonal basis
\begin{align}
\label{eq:cbqqqq}
\theta_{1} = (t^a)_{ij}(t^a)_{kl}\,,
\qquad
\theta_{2} = \delta_{ij}\delta_{kl}\,,
\\
\label{eq:cbdqqqq}
\theta_{1}^{\dagger} = (t^a)_{ji}(t^a)_{lk}\,,
\qquad
\theta_{2}^{\dagger} = \delta_{ji}\delta_{lk}\,.
\end{align}
We note that other bases have been used in the literature~\cite{Moult:2015aoa}. We now explicitly perform the calculation for the $qq'\rightarrow qq'$, $qq'\rightarrow q'q$, and $qq\rightarrow qq$ subprocesses. For these subprocesses, we can write
\begin{align}
    \mathcal{M} = \mathcal{M}_{t}^{\rm kin}\left(t^b\right)_{31}\left(t^b\right)_{42}+\mathcal{M}_{u}^{\rm kin}\left(t^b\right)_{32}\left(t^b\right)_{41}\,
    \label{eq:qqqq_upol}
\end{align}
where we have suppressed the $ab\rightarrow cd$ subprocess label. The subscript in the $\mathcal{M}$ terms denotes the relevant Mandelstam variable ($\hat t$ or $\hat u$) for the channel that contributes to the subprocess as shown in the Fig.~\ref{fig:M_upol}. To arrive at this expressions, we have separated the color parts from the kinematic parts (denoted with the superscript $\rm{kin}$). These kinematic scattering amplitudes are defined by 
\begin{align}
\mathcal{M}_t^{\rm kin} = 
\begin{cases} 
    -\dfrac{g_s^2}{\hat{t}}\bar{u}(P_1)\gamma^\mu u(x_a\,P_A)\bar{u}(P_2)\gamma_\mu u(x_b\,P_B) & \hspace{0.2cm}\qquad \quad \hspace{0.1cm} ab\rightarrow cd = qq'\rightarrow qq'\\
    0 & \hspace{0.2cm}\textrm{for} \qquad ab\rightarrow cd = qq'\rightarrow q'q\\
    -\dfrac{g_s^2}{\hat{t}}\bar{u}(P_1)\gamma^\mu u(x_a\,P_A)\bar{u}(P_2)\gamma_\mu u(x_b\,P_B) & \hspace{0.2cm}\qquad \quad \hspace{0.1cm} ab\rightarrow cd = qq\rightarrow qq \,,
\end{cases}
\label{eq:kin_ts}
\end{align}
\begin{align}
\mathcal{M}_u^{\rm kin} = 
\begin{cases} 
    0 & \hspace{0.2cm}\qquad \quad \hspace{0.05cm} ab\rightarrow cd = qq'\rightarrow qq'\\
    -\dfrac{g_s^2}{\hat{u}}\bar{u}(P_2)\gamma^\mu u(x_a\,P_A)\bar{u}(P_1)\gamma_\mu u(x_b\,P_B) & \hspace{0.2cm}\textrm{for} \qquad ab\rightarrow cd = qq'\rightarrow q'q\\
    \dfrac{g_s^2}{\hat{u}}\bar{u}(P_2)\gamma^\mu u(x_a\,P_A)\bar{u}(P_1)\gamma_\mu u(x_b\,P_B) & \hspace{0.2cm}\qquad \quad \hspace{0.05cm} ab\rightarrow cd = qq\rightarrow qq \,.
\end{cases}
\label{eq:kin_us}
\end{align}
We can now decompose these scattering amplitudes in color space as
\begin{align}
    \mathcal{M} = \mathcal{M}_{1}\,\theta_1+\mathcal{M}_{2}\,\theta_2
    \qquad
    \mathcal{M}^{\dagger} = \mathcal{M}_{1}^{\dagger}\,\theta_1^{\dagger}+\mathcal{M}_{2}^{\dagger}\,\theta_2^{\dagger} \,,
\end{align}
where
\begin{align}
    \mathcal{M}_{1} = \frac{\Tr\left[\mathcal{M} \theta_1^{\dagger}\right]}{\Tr\left[\theta_1 \theta_1^{\dagger}\right]}\,
    \qquad
    \mathcal{M}_{2} = \frac{\Tr\left[\mathcal{M} \theta_2^{\dagger}\right]}{\Tr\left[\theta_2 \theta_2^{\dagger}\right]}\,
    \qquad
    \mathcal{M}_{1}^{\dagger} = \frac{\Tr\left[\mathcal{M}^{\dagger} \theta_1\right]}{\Tr\left[\theta_1 \theta_1^{\dagger}\right]}\,
    \qquad
    \mathcal{M}_{2}^{\dagger} = \frac{\Tr\left[\mathcal{M}^{\dagger} \theta_2\right]}{\Tr\left[\theta_2 \theta_2^{\dagger}\right]}\,.
    \label{eq:qqqqcolor}
\end{align}
To obtain the expressions in Eq.~\eqref{eq:qqqqcolor}, we have exploited the orthogonality of our chosen color basis in Eqs.\ \eqref{eq:cbqqqq} and\ \eqref{eq:cbdqqqq}.
Then we will have $|\mathcal{M}|^2$ as
\begin{align}
    |\mathcal{M}|^2 = \Tr\left[
    \bm{H}_{ab\rightarrow cd}\cdot \bm{S}_{ab\rightarrow cd}
    \right]\,,
\end{align}
where the hard matrix is given by
\begin{align}
    \bm{H}_{ab\rightarrow cd} = 
    \begin{bmatrix}
    |\mathcal{M}_{1}|^2 & \mathcal{M}_{1}\mathcal{M}_{2}^{\dagger} \\
    \mathcal{M}_{2}\mathcal{M}_{1}^{\dagger} & |\mathcal{M}_{2}|^2
    \end{bmatrix}\,,
\end{align}
and the leading order soft matrix as
\begin{align}
    \bm{S}_{ab\rightarrow cd} = 
    \begin{bmatrix}
    \Tr\left[\theta_1 \theta_1^{\dagger}\right] & \Tr\left[\theta_1 \theta_2^{\dagger}\right] \\
    \Tr\left[\theta_2 \theta_1^{\dagger}\right] & \Tr\left[\theta_2 \theta_2^{\dagger}\right]
    \end{bmatrix}
     = 
     \begin{bmatrix}
    \frac{1}{2}N_c C_F & 0 \\
    0 & N_c^2
    \end{bmatrix}\,.
\end{align}
The hard matrices of the four quark processes in $\Gamma_{31}\Gamma_{42}$ color basis in Tab.\ \ref{tab:quark-channels} are given by
\begin{align}
    \bm{H}_{qq'\rightarrow qq'} =& \frac{8 g_s^4 \left(\hat{s}^2+\hat{u}^2\right)}{\hat{t}^2}
    \begin{bmatrix}
 1 & 0 \\
 0 & 0 \\
    \end{bmatrix}\,,
\end{align}
\begin{align}
    \bm{H}_{qq'\rightarrow q'q} =& \frac{8 g_s^4 \left(\hat{s}^2+\hat{t}^2\right)}{\hat{u}^2 C_A^2}
\begin{bmatrix}
 1 & -C_F \\
 -C_F & C_F^2 \\
\end{bmatrix}\,,
\end{align}
\begin{align}
    \bm{H}_{qq\rightarrow qq} = \frac{8 g_s^4}{\hat{t}^2 \hat{u}^2 N_c^2}
\begin{bmatrix}
 \hat{t}^4+\hat{s}^2 \hat{t}^2-2 N_c \hat{s}^2 \hat{u} \hat{t}+N_c^2 \hat{u}^4+N_c^2 \hat{s}^2 \hat{u}^2 & -C_F \hat{t} \left(\hat{t}^3+\hat{s}^2 \hat{t}-N_c \hat{s}^2
   \hat{u}\right) \\
 -C_F \hat{t} \left(\hat{t}^3+\hat{s}^2 \hat{t}-N_c \hat{s}^2 \hat{u}\right) & C_F^2 \hat{t}^2 \left(\hat{s}^2+\hat{t}^2\right) \\
\end{bmatrix}\,.
\end{align}
We find these results to be consistent with the expressions in \cite{Kelley:2010fn}. The remaining hard functions can be obtained from crossing symmetries.
\subsubsection{Two quarks and two gluon subprocesses}
\begin{table}[t!]
\def\arraystretch{2.}
\setlength{\tabcolsep}{5pt}
 \begin{tabular}{|| c | c || c | c || c | c || c | c || c | c || c | c ||} 
 \hline
 12 $\rightarrow$ 34 & Basis & 12 $\rightarrow$ 34 & Basis & 12 $\rightarrow$ 34 & Basis & 12 $\rightarrow$ 34 & Basis & 12 $\rightarrow$ 34 & Basis & 12 $\rightarrow$ 34 & Basis \\ 
 \hline
 \hline
 $q\bar{q}\rightarrow gg$ & $\Gamma_{n,21}^{ab}$ & 
 $qg\rightarrow gq$       & $\Gamma_{n,41}^{ab}$ & 
 $qg\rightarrow qg$       & $\Gamma_{n,31}^{ab}$ & 
 $gq\rightarrow gq$       & $\Gamma_{n,42}^{ab}$ & 
 $gq\rightarrow qg$       & $\Gamma_{n,32}^{ab}$ & 
 $gg\rightarrow q\bar{q}$ & $\Gamma_{n,43}^{ab}$ \\
\hline
 $\bar{q}g\rightarrow \bar{q}g$ & $\Gamma_{n,21}^{ab}$ & 
 $\bar{q}g\rightarrow g\bar{q}$ & $\Gamma_{n,41}^{ab}$ & 
 $\bar{q}q\rightarrow gg$       & $\Gamma_{n,31}^{ab}$ & 
 $gg\rightarrow \bar{q}q$       & $\Gamma_{n,42}^{ab}$ & 
 $g\bar{q}\rightarrow \bar{q}g$ & $\Gamma_{n,32}^{ab}$ & 
 $g\bar{q}\rightarrow g\bar{q}$ & $\Gamma_{n,43}^{ab}$ \\
\hline
\end{tabular}
    \caption{The choice of basis for each of two quark two gluon subprocesses. Three operators $\Gamma_{1,ij}^{ab}$,$\Gamma_{2,ij}^{ab}$,$\Gamma_{3,ij}^{ab}$ are required to span the color space for each subprocess.}
    \label{tab:quark-gluon-channels}
\end{table}
In Tab.~\ref{tab:quark-gluon-channels}, we provide a list of subprocesses involving two quarks and two gluons with the color basis operators $\Gamma_{n,ij}^{ab}$. For the two quark and two gluon interactions, three operators, $n=1,2,3$, are required to span the color space. A convenient choice for the computation is the set of orthogonal operators (primed),
\begin{align}
    {\Gamma^{ab}_{1, ij}}' = \frac{\delta^{ab}}{2 N_c} \delta_{ij}\,,
    \qquad
    {\Gamma^{ab}_{2, ij}}' = \frac{1}{2}d^{abc}t^{c}_{ij}\,,
    \qquad
    {\Gamma^{ab}_{3, ij}}' = \frac{1}{2}f^{abc}t^{c}_{ij}\,,
\end{align}
which has the corresponding orthogonal basis,
\begin{align}
    \theta_1' = \frac{\delta^{ab}}{2 N_c} \delta_{ij}\,,
    \qquad
    \theta_2' = \frac{1}{2}d^{abc}t^{c}_{ij}\,,
    \qquad
    \theta_3' = \frac{1}{2}f^{abc}t^{c}_{ij}\,.
    \label{eq:primed}
\end{align}
At the same time, we find that the final expressions for the hard matrices take a simpler form when one uses the non-orthogonal basis used in Refs.~\cite{Liu:2014oog,Kelley:2010fn,Moult:2015aoa} by defining the basis operators to be (unprimed)
\begin{align}
    \Gamma^{ab}_{1, ij} = (t^a t^b)_{ij}\,,
    \qquad
    \Gamma^{ab}_{2, ij} = (t^b t^a)_{ij}\,,
    \qquad
    \Gamma^{ab}_{3, ij} = \delta_{ij} \delta^{ab}\,.
\end{align}
The corresponding basis is given by
\begin{align}
    \theta_1 = (t^a t^b)_{ij}\,,
    \qquad
    \theta_2 = (t^b t^a)_{ij}\,,
    \qquad
    \theta_3 = \delta_{ij} \delta^{ab}\,.
    \label{eq:unprimed}
\end{align}
We note that the normalization of $\theta_{3}$ in \cite{Moult:2015aoa} differs from the normalization of  Refs.~\cite{Liu:2014oog,Kelley:2010fn} by a factor of 2. 
For the choice of basis in Eq.\ \eqref{eq:unprimed}, the LO soft matrix is given by 
\begin{align}
    \bm{S}_{ab\rightarrow cd} = 
    \begin{bmatrix}
    \Tr\left[\theta_1 \theta_1^{\dagger}\right] & \Tr\left[\theta_1 \theta_2^{\dagger}\right] & \Tr\left[\theta_1 \theta_3^{\dagger}\right]\\
    \Tr\left[\theta_2 \theta_1^{\dagger}\right] & \Tr\left[\theta_2 \theta_2^{\dagger}\right] & \Tr\left[\theta_2 \theta_3^{\dagger}\right]\\
    \Tr\left[\theta_3 \theta_1^{\dagger}\right] & \Tr\left[\theta_3 \theta_2^{\dagger}\right] & \Tr\left[\theta_3 \theta_3^{\dagger}\right]
    \end{bmatrix}
    =
    \begin{bmatrix}
     N_c C_F^2 & -\frac{C_F}{2} & N_c C_F \\  
     -\frac{C_F}{2} & N_c C_F^2 & N_c C_F \\
     N_c C_F & N_c C_F & 2N_c^2 C_F \\
    \end{bmatrix}\,.
\end{align}
In order to exploit the orthogonality condition of the primed basis in Eq.\ \eqref{eq:primed}, but still provide a simple expression for the hard matrices using the unprimed basis in Eq.\ \eqref{eq:unprimed}, we first compute the hard matrices in the primed basis then obtain the results in the unprimed basis using the relation
\begin{align}
    \bm{H}_{ab\rightarrow cd} = \bm{R}^{\dagger}\, \bm{H}_{ab\rightarrow cd}'\, \bm{R}\,,
    \qquad\text{where}\quad
    \bm{R} = 
    \begin{bmatrix}
    1 & 1 & -1\\
    1 & 1 &  1\\
    2 N_c & 0 & 0
    \end{bmatrix}^{-1}\,.
    \label{eq:Hrot}
\end{align}
We now perform the calculation for the hard matrices for the $q\bar{q}\rightarrow gg$ process in the primed orthogonal basis. The scattering amplitude for this subprocess can be written in color space as
\begin{align}
    \mathcal{M} = M_{1}\theta_1'+M_{2}\theta_2'+M_{3}\theta_3'
    \qquad
    \mathcal{M}^{\dagger} = M_{1}^{\dagger}{\theta_1'}^{\dagger}+M_{2}^{\dagger}{\theta_2'}^{\dagger}+M_{3}^{\dagger}{\theta_3'}^{\dagger}
\end{align}
where 
\begin{align}
    \mathcal{M}_{1} = \frac{\Tr\left[\mathcal{M} {\theta_1'}^{\dagger}\right]}{\Tr\left[{\theta_1'} {\theta_1'}^{\dagger}\right]}\,
    \qquad
    \mathcal{M}_{2} = \frac{\Tr\left[\mathcal{M} {\theta_2'}^{\dagger}\right]}{\Tr\left[{\theta_2'} {\theta_2'}^{\dagger}\right]}\,,
    \qquad
    \mathcal{M}_{3} = \frac{\Tr\left[\mathcal{M} {\theta_3'}^{\dagger}\right]}{\Tr\left[{\theta_3'} {\theta_3'}^{\dagger}\right]}\,,
    \\
    \mathcal{M}_{1}^{\dagger} = \frac{\Tr\left[\mathcal{M}^{\dagger} {\theta_1'}\right]}{\Tr\left[{\theta_1'} {\theta_1'}^{\dagger}\right]}\,
    \qquad
    \mathcal{M}_{2}^{\dagger} = \frac{\Tr\left[\mathcal{M}^{\dagger} {\theta_2'}\right]}{\Tr\left[{\theta_2'} {\theta_2'}^{\dagger}\right]}\,,
    \qquad
    \mathcal{M}_{3}^{\dagger} = \frac{\Tr\left[\mathcal{M}^{\dagger} {\theta_3'}\right]}{\Tr\left[{\theta_3'} {\theta_3'}^{\dagger}\right]}\,.
    \label{eq:qqbggcolor}
\end{align}
The hard matrix in the primed basis can therefore be computed as
\begin{align}
    \bm{H}'_{q\bar{q}\rightarrow gg} =
    \begin{bmatrix}
    M_{1}\,M_{1}^{\dagger} & M_{1}\,M_{2}^{\dagger} & M_{1}\,M_{3}^{\dagger}\\
    M_{2}\,M_{1}^{\dagger} & M_{2}\,M_{2}^{\dagger} & M_{2}\,M_{3}^{\dagger}\\
    M_{3}\,M_{1}^{\dagger} & M_{3}\,M_{2}^{\dagger} & M_{3}\,M_{3}^{\dagger}\\
    \end{bmatrix}\,.
\end{align}
Finally, we now use Eq.~\eqref{eq:Hrot} to obtain the simplified hard functions in the unprimed basis  as
\begin{align}
    \bm{H}_{q\bar{q}\rightarrow gg} = 8 g_s^4 \frac{\left(\hat{t}^2+\hat{u}^2\right)}{\hat{s}^2}
    \begin{bmatrix}
 \frac{\hat{u}}{\hat{t}} & 1 & 0 \\
 1 & \frac{\hat{t}}{\hat{u}} & 0 \\
 0 & 0 & 0 \\
    \end{bmatrix}\,.
\end{align}
The hard matrices for other subprocesses involving two quarks and two gluons, such as $qg\to qg$, can be obtained from this expression using crossing symmetries.
\subsubsection{Four gluon subprocesses}
For the four gluon subprocesses, $gg\to gg$, we follow the work in Refs.~\cite{Kelley:2010fn,Liu:2014oog} to use the following over-complete basis
\begin{alignat}{3}
    &\theta_1 = \Tr\left[t^{a_1}t^{a_2}t^{a_3} t^{a_4}\right]\,,
    \qquad
    &&\theta_2 = \Tr\left[t^{a_1} t^{a_2} t^{a_4} t^{a_3}\right]\,,
    \qquad
    & &\theta_3 = \Tr\left[t^{a_1} t^{a_4} t^{a_3} t^{a_2}\right]\,,
\notag\\
    &\theta_4 = \Tr\left[t^{a_1} t^{a_4} t^{a_2} t^{a_3}\right]\,,
    \qquad
    &&\theta_5 = \Tr\left[t^{a_1} t^{a_3} t^{a_4} t^{a_2}\right]\,,
    \qquad
    & &\theta_6 = \Tr\left[t^{a_1} t^{a_3} t^{a_2} t^{a_4}\right]\,,
\notag\\
    &\theta_7 = \Tr\left[t^{a_1} t^{a_4}\right]\Tr\left[ t^{ a_2} t^{a_3}\right]\,,
    \qquad
    &&\theta_8 = \Tr\left[t^{a_1} t^{a_2}\right]\Tr\left[t^{a_3} t^{a_4}\right]\,,
    \qquad
    & &\theta_9 = \Tr\left[t^{a_1} t^{a_3}\right]\Tr\left[t^{a_2} t^{a_4}\right]\,.
    \label{eq:4-glu-upol}
\end{alignat}
We note that a six dimensional basis was chosen in~\cite{Moult:2015aoa}. Using this basis in Eq.~\eqref{eq:4-glu-upol}, one can show that the hard matrix takes the following form
\begin{align}
    \bm{H}_{gg\to gg} = \frac{2 g_s^4 \left(\hat{s}^4+\hat{t}^4+\hat{u}^4\right)}{\hat{s}^2 \hat{u}^2 N_c^2 C_F^2}
    \begin{bmatrix}
 1 & \frac{\hat{u}}{\hat{t}} & 1 & \frac{\hat{s}}{\hat{t}} & \frac{\hat{u}}{\hat{t}} & \frac{\hat{s}}{\hat{t}} & 0 & 0 & 0 \\
 \frac{\hat{u}}{\hat{t}} & \frac{\hat{u}^2}{\hat{t}^2} & \frac{\hat{u}}{\hat{t}} & \frac{\hat{s} \hat{u}}{\hat{t}^2} & \frac{\hat{u}^2}{\hat{t}^2} & \frac{\hat{s} \hat{u}}{\hat{t}^2} & 0 & 0 & 0 \\
 1 & \frac{\hat{u}}{\hat{t}} & 1 & \frac{\hat{s}}{\hat{t}} & \frac{\hat{u}}{\hat{t}} & \frac{\hat{s}}{\hat{t}} & 0 & 0 & 0 \\
 \frac{\hat{s}}{\hat{t}} & \frac{\hat{s} \hat{u}}{\hat{t}^2} & \frac{\hat{s}}{\hat{t}} & \frac{\hat{s}^2}{\hat{t}^2} & \frac{\hat{s} \hat{u}}{\hat{t}^2} & \frac{\hat{s}^2}{\hat{t}^2} & 0 & 0 & 0 \\
 \frac{\hat{u}}{\hat{t}} & \frac{\hat{u}^2}{\hat{t}^2} & \frac{\hat{u}}{\hat{t}} & \frac{\hat{s} \hat{u}}{\hat{t}^2} & \frac{\hat{u}^2}{\hat{t}^2} & \frac{\hat{s} \hat{u}}{\hat{t}^2} & 0 & 0 & 0 \\
 \frac{\hat{s}}{\hat{t}} & \frac{\hat{s} \hat{u}}{\hat{t}^2} & \frac{\hat{s}}{\hat{t}} & \frac{\hat{s}^2}{\hat{t}^2} & \frac{\hat{s} \hat{u}}{\hat{t}^2} & \frac{\hat{s}^2}{\hat{t}^2} & 0 & 0 & 0 \\
 0 & 0 & 0 & 0 & 0 & 0 & 0 & 0 & 0 \\
 0 & 0 & 0 & 0 & 0 & 0 & 0 & 0 & 0 \\
 0 & 0 & 0 & 0 & 0 & 0 & 0 & 0 & 0 \\
    \end{bmatrix}\,.
\end{align}
The LO soft matrix for this channel is given in Appendix C of \cite{Liu:2014oog} for this basis as
\begin{align}
    \bm{S}_{gg\to gg} = \frac{C_F}{8 N_c}
    \begin{bmatrix}
 a_0 & b_0 & c_0 & b_0 & b_0 & b_0 & d_0     & d_0 &-e_0 \\
 b_0 & a_0 & b_0 & b_0 & c_0 & b_0 &-e_0     & d_0 & b_0 \\
 c_0 & b_0 & a_0 & b_0 & b_0 & b_0 & d_0     & d_0 &-e_0 \\
 b_0 & b_0 & b_0 & a_0 & b_0 & c_0 & d_0     &-e_0 & d_0 \\
 b_0 & c_0 & b_0 & b_0 & a_0 & b_0 &-e_0     & d_0 & d_0 \\
 b_0 & b_0 & b_0 & c_0 & b_0 & a_0 & d_0     &-e_0 & d_0 \\
 d_0 &-e_0 & d_0 & d_0 &-e_0 & d_0 & d_0 e_0 & e_0^2 & e_0^2 \\
 d_0 & d_0 & d_0 &-e_0 & d_0 &-e_0 & e_0^2   & d_0 e_0 & e_0^2 \\
-e_0 & d_0 &-e_0 & d_0 & d_0 & d_0 & e_0^2   & e_0^2 & d_0 e_0 \\
    \end{bmatrix}\,,
\end{align}
where $a_0 = N_c^4-3 N_c^2+3$, $b_0 = 3-N_c^2$, $c_0 = 3+N_c^2$, $d_0 = 2 N_c^2 C_F$, and $e_0 = N_c$.

\subsection{Polarized Hard Matrices}
As we have emphasized in the previous section, Sivers function is non-universal. The well-known example is the sign change between the Sivers function probed in SIDIS and that in Drell-Yan (DY) process~\cite{Brodsky:2002rv,Collins:2002kn,Boer:2003cm},
\bea
 f_{1T}^{\perp\,q {\rm (DY)}}(x, k_\perp, \mu) = - f_{1T}^{\perp\,q {\rm (SIDIS)}}(x, k_\perp, \mu) \,.
 \label{eq:sign}
\eea
Such a sign change can be easily taken care of in describing the Drell-Yan Sivers asymmetry, 
\bea
d\Delta\sigma(S_\perp) \propto f_{1T}^{\perp\,q {\rm (DY)}}(x, k_\perp, \mu) H(Q, \mu)  = f_{1T}^{\perp\,q {\rm (SIDIS)}}(x, k_\perp, \mu) \big[- H(Q, \mu)\big]\,,
\label{eq:sign-hard}
\eea
where $H(Q, \mu)$ is the hard function in the Drell-Yan process, and we have applied Eq.~\eqref{eq:sign} in the second step. In other words, if we use the SIDIS Sivers function in a Drell-Yan process, we shift the minus sign (or the process-dependence) into the hard function. 

For the partonic subprocesses in the hadronic dijet production, one has much more complicated process-dependence for the Sivers functions involved. This can be seen from the highly nontrivial gauge link structure which has been derived in~\cite{Bomhof:2006dp} in the definition of the TMD PDFs. Even in these complicated processes, one can incorporate such process-dependence of the Sivers functions into modified hard functions as in Eq.\ \eqref{eq:sign-hard}~\cite{Bacchetta:2005rm,Bomhof:2007su,Qiu:2007ey,Vogelsang:2007jk,Qiu:2007ar}. We follow a similar procedure in this section to include this process-dependence of the Sivers functions into the hard functions in the matrix form. 

\begin{figure}
  \centering
  \includegraphics[valign = c,scale = 0.45]{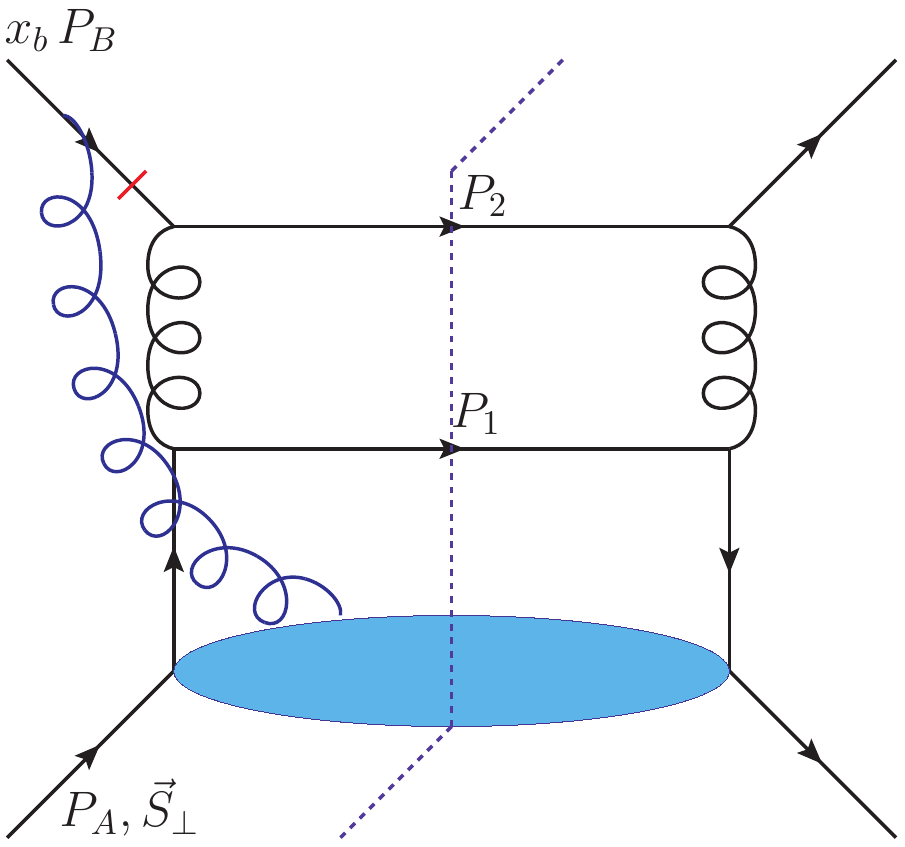} = 
  \includegraphics[valign = c,scale = 0.45]{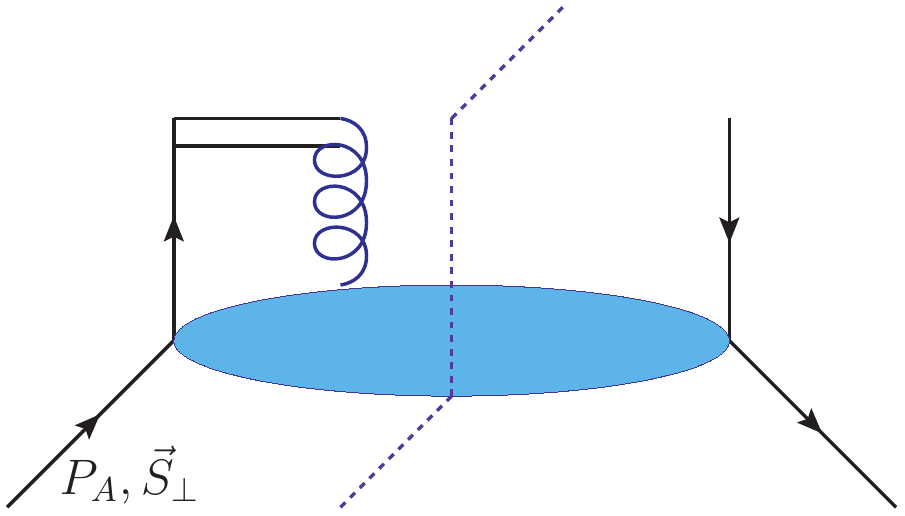}
  \includegraphics[valign = c,scale = 0.45]{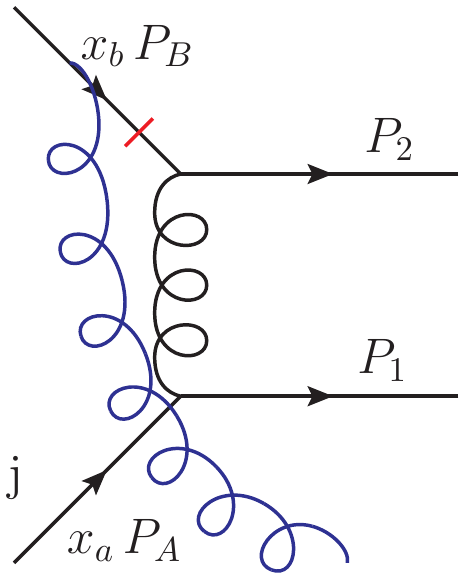}
  \includegraphics[valign = c,scale = 0.45]{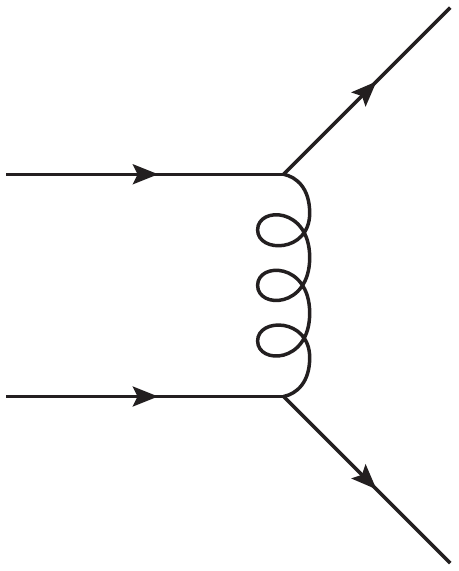} $\times \dfrac{t^{a}_{j1}}{C_F}$
  \includegraphics[valign = c,scale = 0.45]{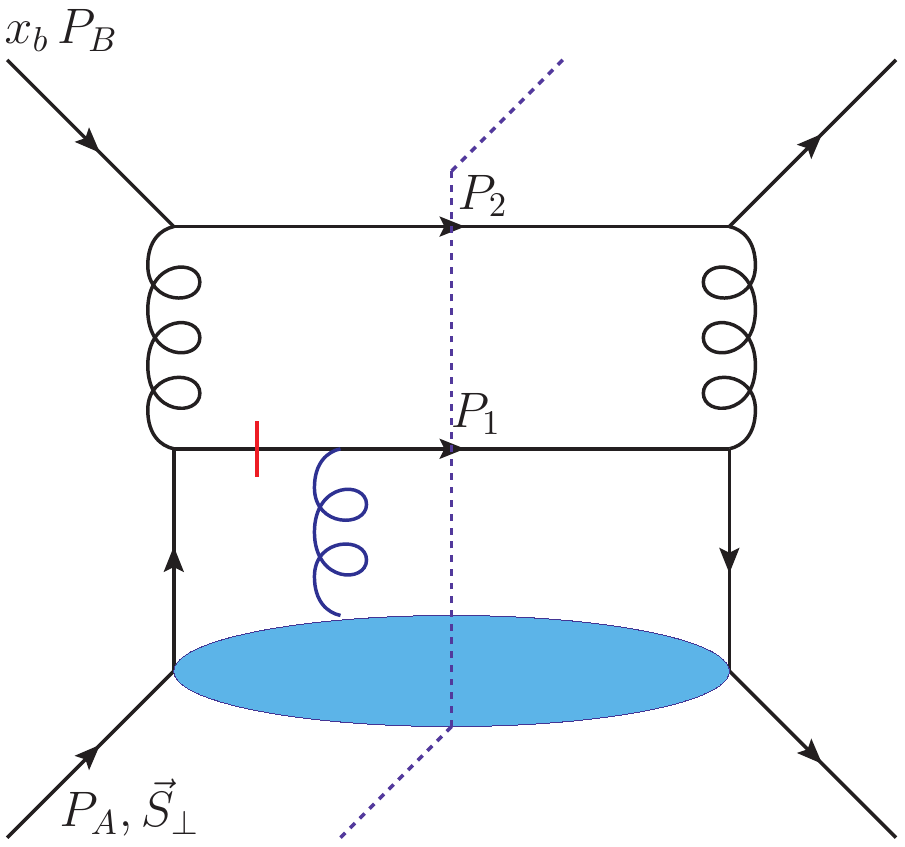} = 
  \includegraphics[valign = c,scale = 0.45]{\FigPath/Sivers_dijet-cropped.pdf}
  \includegraphics[valign = c,scale = 0.45]{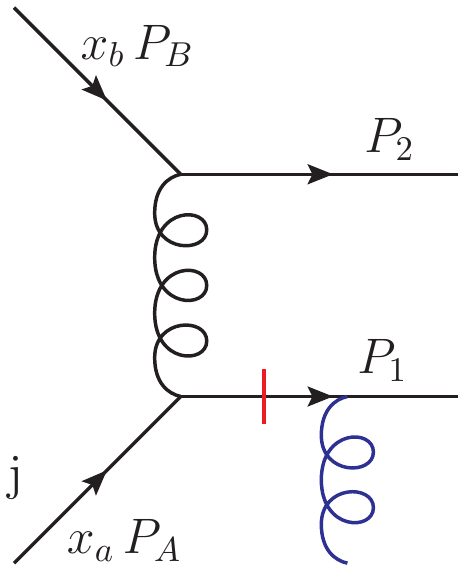}
  \includegraphics[valign = c,scale = 0.45]{\FigPath/CI-hard-upol-cropped.pdf} $\times \dfrac{t^{a}_{j1}}{C_F}$
  \includegraphics[valign = c,scale = 0.45]{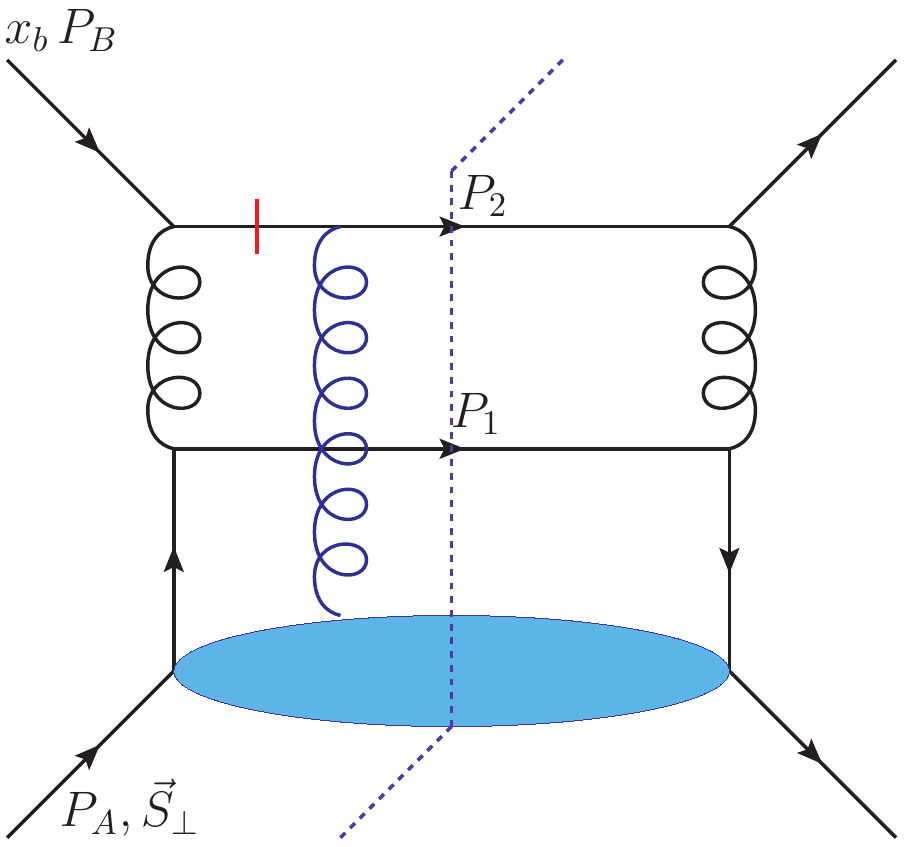} = 
  \includegraphics[valign = c,scale = 0.45]{\FigPath/Sivers_dijet-cropped.pdf}
  \includegraphics[valign = c,scale = 0.45]{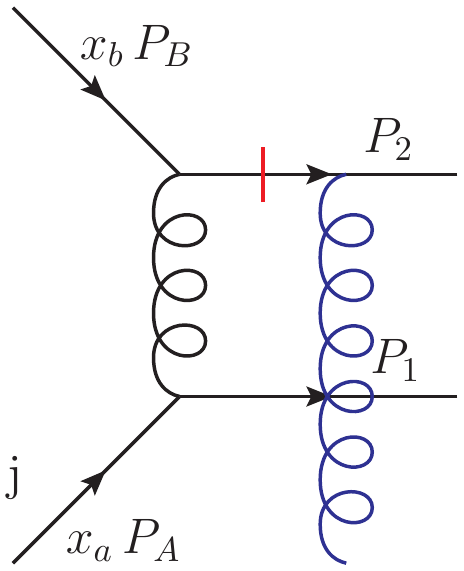}
  \includegraphics[valign = c,scale = 0.45]{\FigPath/CI-hard-upol-cropped.pdf} $\times \dfrac{t^{a}_{j1}}{C_F}$
  \caption{A demonstration of the factorization between the Sivers function and the hard function for $qq'\to qq'$ subprocess. The red lines indicate the locations of the soft poles while the blue gluon represents the gauge link which generates the asymmetry.}
  \label{fig:Factorization_pol}
\end{figure}

In Fig.~\ref{fig:Factorization_pol}, we demonstrate the factorization between the Sivers function and modified hard functions. Unlike the unpolarized case, the contributions of the Sivers asymmetry are given by considering the attachment of an additional collinear (to the incoming hadron) gluon to three of the external legs. Such a gluon is part of the gauge link in the definition of the Sivers function, and it is the imaginary part of the Feynman diagram (related to the so-called soft gluonic pole) that contributes to the process-dependence of the Sivers function.   

It is important to note that the additional gluon leads to additional complications so that naive crossing symmetry cannot be used to relate one hard function to another, as in the unpolarized case studied above. These complications occur because the contributions to the Sivers asymmetry are only given by attaching the additional gluon to three of the four external legs. Furthermore, since the sign of the interaction (imaginary part) with the external gluon is opposite for quarks and anti-quarks, this sign must also be accounted for when applying crossing symmetry or charge conjugation.
\subsubsection{Four quark subprocesses}
\begin{figure}
  \centering
  \includegraphics[valign = b, height = 1.in]{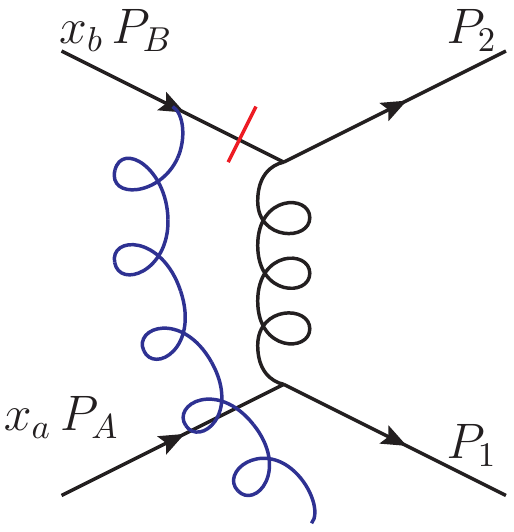}
  \qquad
  \includegraphics[valign = b, height = 1.in]{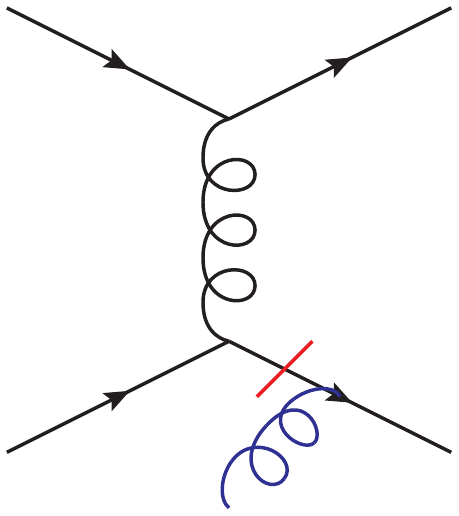} 
  \qquad
  \includegraphics[valign = b, height = 1.in]{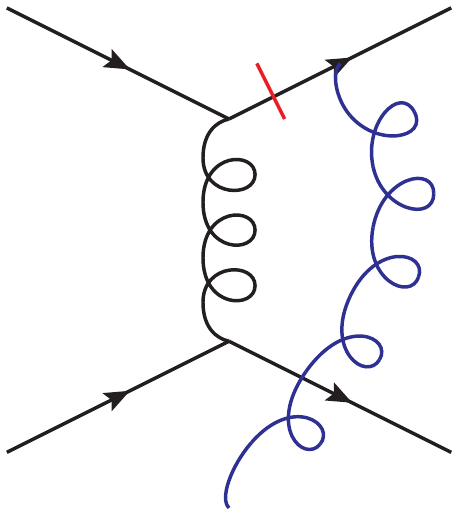}
  \qquad
  \includegraphics[valign = b, height = 1.in]{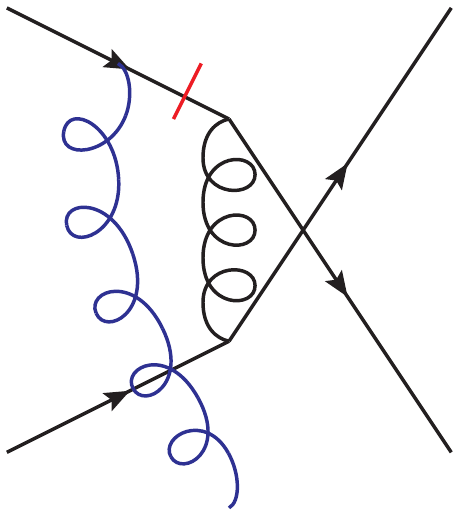} 
  \qquad 
  \includegraphics[valign = b, height = 1.in]{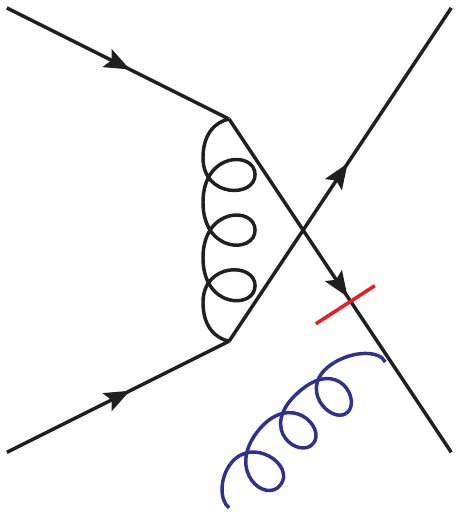}
  \qquad 
  \includegraphics[valign = b, height = 1.in]{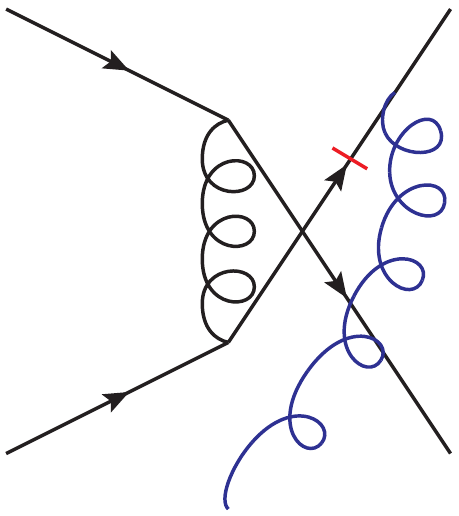}
  \caption{Polarized scattering amplitudes for the $qq\rightarrow qq$ subprocess. From left the right, the first three graphs give the scattering amplitude for the $t$-channel for initial-state, final-state 1, and final-state 2 interactions. The remaining channels give the contributions for the $u$-channel for initial-state, final-state 1, and final-state 2 interactions.}
  \label{fig:MSiversj}
\end{figure}
As in the unpolarized case, the bases for four quark subprocesses are given in Tab.~\ref{tab:quark-channels}. As discussed above, 
one cannot naively apply crossing symmetry to obtain hard matrices of a general polarized subprocess. For the polarized four quark subprocesses, however, only the sign of each color factor changes under charge conjugation. Therefore, the hard matrices for the bottom row of Tab.~\ref{tab:quark-channels} can be obtained from the results from the top row of this table with the addition of a minus sign. 

To demonstrate how $\bm{H}^{\rm Sivers}_{ab\rightarrow cd}$ are derived, we explicitly perform the calculation for the $qq'\rightarrow qq'$, $qq'\rightarrow q'q$, and $qq\rightarrow qq$ subprocesses as we did for the unpolarized case. Afterwards, we provide the expressions for the remaining subprocesses. To start, it is important to remind ourselves that a non-vanishing Sivers asymmetry requires initial/final state interactions generating a phase. Because all initial and final partonic states relevant for dijet production are colored, both initial and final state interactions have to be taken into account. Such interactions would generate non-trivial gauge link structures, see e.g. Refs.~\cite{Bomhof:2004aw,Bomhof:2006dp,Vogelsang:2007jk}. On the left side of Fig.~\ref{fig:Factorization_pol}, as an example, we show all possible diagrams with one gluon exchange between the remnant of the polarized proton and the $qq'\rightarrow qq'$ hard scattering part, which contribute to the Sivers asymmetry. Now with the presence of the extra gluon scattering (first order of the gauge link expansion), the diagram at the left side of the cut will be denoted as $\mathcal{M}^{\textrm{Sivers}, a}_{j}$, while the right side is same as the unpolarized case denoted as $\mathcal{M}^{\dagger}$. Here $a$ is the color for the attached gluon, $j$ is the color index for the incoming quark with momentum $x_a P_A$ on the left side of cut line, while the color index for the incoming quark on the right side of the cut line is given by $1$ like in the previous section. In contrast to the unpolarized correlation function, quarks $j$ and $1$ do not need to have the same color, because of the presence of the gluon from the gauge link. Now we perform the following expansion to obtain the hard matrix $|\mathcal{M}^{\rm Sivers}|^2$ for the polarized case,
\bea
\mathcal{M}^{\textrm{Sivers}, a}_{j} \mathcal{M}^{\dagger} = |\mathcal{M}^{\rm Sivers}|^2\, t^a_{1j}\,,
\label{e.Mj1}
\eea
where $t^a_{1j}$ will be included into the quark-quark correlator in the polarized proton to  become $\sim \langle PS|\bar\psi_1\, n\cdot A^a t^a_{1j}\, \psi_j |PS \rangle$, see e.g. Ref.~\cite{Bomhof:2007zz,Bomhof:2006dp,Qiu:2007ey}. From Eq.~\eqref{e.Mj1}, we thus derive
\bea
|\mathcal{M}^{\rm Sivers}|^2 =& \frac{1}{\textrm {Tr}\left[t^a t^a\right]} \mathcal{M}^{\textrm{Sivers}, a}_{j}\, t^{a}_{j1}\, \mathcal{M}^{\dagger} 
\nn \\
=& \frac{1}{N_c}\cdot \frac{1}{C_F} \mathcal{M}^{\textrm{Sivers}, a}_{j}\, t^{a}_{j1}\, \mathcal{M}^{\dagger}\,.
\label{e.M_siv_0}
\eea
At the same time, we use the convention that $N_{\rm init}$ in the polarized and unpolarized cases are the same. Therefore, the factor of $1/N_c$ in Eq.~\eqref{e.M_siv_0} is absorbed into $N_{\rm init}$. With that in mind, to arrive at the correct normalization of the polarized hard function, we thus obtain
\bea
|\mathcal{M}^{\rm Sivers}|^2 \to \frac{1}{C_F} \mathcal{M}^{\textrm{Sivers}, a}_{j} \, t^{a}_{j1}\, \mathcal{M}^{\dagger}\,,
\label{e.M_siv}
\eea
which is demonstrated on the right-hand side of Fig.~\ref{fig:Factorization_pol}.

Now we need to project $\mathcal{M}^{\textrm{Sivers}, a}_{j}$ and $\mathcal{M}^{\dagger}$ into the color basis separately. The polarized scattering amplitude $\mathcal{M}^{\textrm{Sivers}, a}_{j}$ can be written as 
\begin{align}
\label{eq:MSiversj}
    \mathcal{M}_{j}^{\textrm{Sivers}, a} &= \mathcal{M}^{\rm kin}_t \left(t^b t^a\right)_{42}\left(t^b \right)_{3j}+\mathcal{M}^{\rm kin}_t \left(t^b \right)_{42}\left(t^a t^b \right)_{3j}+\mathcal{M}^{\rm kin}_t \left(t^a t^b\right)_{42}\left(t^b \right)_{3j}\\
    & +\mathcal{M}^{\rm kin}_u \left(t^b t^a\right)_{32}\left(t^b \right)_{4j}+\mathcal{M}^{\rm kin}_u \left(t^a t^b\right)_{32}\left(t^b \right)_{4j}+\mathcal{M}^{\rm kin}_u \left(t^b \right)_{32}\left(t^a t^b \right)_{4j}\,, \nn
\end{align}
where $\mathcal{M}^{\rm kin}_t$ and $\mathcal{M}^{\rm kin}_u$ are the same as the expressions in Eqs.~\eqref{eq:kin_ts} and \eqref{eq:kin_us}. From left to right on the top line of this expression, these terms give the scattering amplitudes for the initial-state, final-state 1, and final-state 2 interaction for the $t$-channel, corresponding to the first three diagrams of Fig.\ \ref{fig:MSiversj} in the same order. Likewise from left to right on the bottom line, the terms give the scattering amplitude for the initial-state, final-state 1, and final-state 2 interaction for the $u$-channel, corresponding to the last three diagrams of Fig.\ \ref{fig:MSiversj} in the same order. Using the Feynman rules for the gauge link color factors given in Fig.~6 of \cite{Qiu:2007ey}, we easily arrive at Eq.\ \eqref{eq:MSiversj} from these diagrams.
From the unpolarized scattering amplitude given in Eq.\ \eqref{eq:qqqq_upol}, we write the conjugate amplitude as
\begin{align}
    \mathcal{M}^{\dagger} = \mathcal{M}^{\rm kin \dagger}_t \left(t^b \right)_{24}\left(t^b \right)_{13}+\mathcal{M}^{\rm kin \dagger}_u \left(t^b \right)_{23}\left(t^b \right)_{14}\,.
\end{align}
Analogous to the unpolarized scattering amplitude, the scattering amplitude can be decomposed into the orthogonal basis given in Eq.\ \eqref{eq:cbqqqq} as
\begin{align}
    \mathcal{M}_{j}^{\textrm{Sivers}, a}t^{a}_{j1} =& \mathcal{M}_{1}^{\rm Sivers}\theta_{1}
    +\mathcal{M}_{2}^{\rm Sivers}\theta_{2}\,,
\\
    \mathcal{M}_{i}^{\dagger} =& \mathcal{M}_1^{\dagger}\theta_{1}^{\dagger}
    +\mathcal{M}_2^{\dagger}\theta_{2}^{\dagger}\,,
\end{align}
where we have
\begin{align}
    \mathcal{M}_1^{\rm Sivers} =& \frac{\Tr\left[\mathcal{M}_{j}^{\textrm{Sivers}, a} t^{a}_{j1} \, \theta_{1}^{\dagger}\right]}{\Tr\left[\theta_{1}\,\theta_{1}^{\dagger}\right]}\,,
    \qquad
    \mathcal{M}_2^{\rm Sivers} =\frac{\Tr\left[\mathcal{M}_{j}^{\textrm{Sivers}, a} t^{a}_{j1} \, \theta_{2}^{\dagger}\right]}{\Tr\left[\theta_{2}\,\theta_{2}^{\dagger}\right]}\,,
\\
    \mathcal{M}_1^{\dagger} =& \frac{\Tr\left[\mathcal{M}^{\dagger}\, \theta_{1}\right]}{\Tr\left[\theta_{1}\,\theta_{1}^{\dagger}\right]}\,,
    \qquad
    \mathcal{M}_2^{\dagger} =\frac{\Tr\left[\mathcal{M}^{\dagger} \, \theta_{2}\right]}{\Tr\left[\theta_{2}\,\theta_{2}^{\dagger}\right]}\,.
\end{align}
After performing this decomposition, we can now write
\begin{align}
    |\mathcal{M}^{\rm Sivers}|^2 = \Tr\left[\bm{H}^{\rm Sivers}_{ab\rightarrow cd}\cdot \bm{S}_{ab\rightarrow cd}\right]
    \,,
\end{align}
where $\bm{H}^{\rm Sivers}_{ab\rightarrow cd}$ is given by
\begin{align}
    \bm{H}^{\rm Sivers}_{ab\rightarrow cd} = \frac{1}{C_F}
    \begin{bmatrix}
    \mathcal{M}_1^{\rm Sivers}\,M_{1}^{\dagger} & \mathcal{M}_1^{\rm Sivers}\,M_{2}^{\dagger}\\
    \mathcal{M}_2^{\rm Sivers}\,M_{1}^{\dagger} & \mathcal{M}_2^{\rm Sivers}\,M_{2}^{\dagger}\\
    \end{bmatrix}\,
\end{align}
and $\bm{S}$ is the same as the unpolarized case.

From these expressions, we can obtain the polarized hard matrices for the $qq'\rightarrow qq'$, $qq'\rightarrow q'q$, and $qq\rightarrow qq$ subprocesses as
\begin{align}
    \bm{H}^{\rm Sivers}_{qq'\rightarrow qq'} =& \frac{4 g_s^4 \left(\hat{s}^2+\hat{u}^2\right)}{\hat{t}^2 N_c C_F}
\begin{bmatrix}
 N_c^2-5 & 0 \\
 2 C_F & 0 \\
\end{bmatrix}\,,
\\
    \bm{H}^{\rm Sivers}_{qq'\rightarrow q'q} =& -\frac{4 g_s^4 \left(\hat{s}^2+\hat{t}^2\right)}{\hat{u}^2 N_c^3 C_F}
\begin{bmatrix}
 N_c^2+3 & -\left(N_c^2+3\right) C_F \\
 -\left(3-N_c^2\right) C_F & \left(3-N_c^2\right) C_F^2 \\
\end{bmatrix}\,,
\\
    \bm{H}^{\rm Sivers}_{qq\rightarrow qq} =&
    \bm{H}^{\rm Sivers}_{qq'\rightarrow qq'}+\bm{H}^{\rm Sivers}_{qq'\rightarrow q'q} 
    +\frac{4 \hat{s}^2 g_s^4}{\hat{t} \hat{u} N_c^2 C_F}
\begin{bmatrix}
 8 & -\left(5-N_c^2\right) C_F \\
 -\left(5-N_c^2\right) C_F & 2 C_F^2 \\
\end{bmatrix}\,.
\end{align}
Since $qq\to qq$ subprocess receives contributions from both $t$- and $u$-channels (as well as their interference), its expression is the most complicated among the three subprocesses computed. 
One can show that after performing the trace with the soft color matrix, the expressions are consistent with the squared amplitude of \cite{Qiu:2007ey}. The color matrices for the remaining four quark subprocesses in the top row of Tab.\ \ref{tab:quark-channels} can be computed in the same spirit and we obtain the following expressions
\begin{align}
    \bm{H}^{\rm Sivers}_{q\bar{q}\rightarrow q'\bar{q}'} =&\,\frac{4 \left(N_c^2+1\right) g_s^4 \left(\hat{t}^2+\hat{u}^2\right)}{\hat{s}^2 N_c C_F}
\begin{bmatrix}
 1 & 0 \\
 0 & 0 \\
\end{bmatrix}\,,
\\
    \bm{H}^{\rm Sivers}_{q\bar{q}'\rightarrow q\bar{q}'} =&\frac{4 g_s^4 \left(\hat{s}^2+\hat{u}^2\right)}{\hat{t}^2 N_c^3 C_F}
\begin{bmatrix}
 N_c^2+1 & -\left(N_c^2+1\right) C_F \\
 2 N_c C_F^2 & -2 N_c C_F^3 \\
\end{bmatrix}\,,
\\
    \bm{H}^{\rm Sivers}_{q\bar{q}\rightarrow q\bar{q}} =&\,\bm{H}^{\rm Sivers}_{q\bar{q}\rightarrow q'\bar{q}'}+\bm{H}^{\rm Sivers}_{q\bar{q}'\rightarrow q\bar{q}'}-\frac{8 \hat{u}^2 g_s^4}{\hat{s} \hat{t} N_c^2 C_F^2}
\begin{bmatrix}
 \left(N_c^2+1\right) C_F & -\frac{1}{2} \left(N_c^2+1\right) C_F^2 \\
 N_c C_F^3 & 0 \\
\end{bmatrix}\,,
\\
    \bm{H}^{\rm Sivers}_{q\bar{q}'\rightarrow \bar{q}'q} =&-\frac{4 g_s^4 \left(\hat{s}^2+\hat{t}^2\right)}{\hat{u}^2 N_c C_F}
\begin{bmatrix}
 N_c^2-3 & 0 \\
 2 C_F & 0 \\
\end{bmatrix}\,,
\\
    \bm{H}^{\rm Sivers}_{q\bar{q}\rightarrow \bar{q}'q'} =&\,\frac{4 \left(N_c^2+1\right) g_s^4 \left(\hat{t}^2+\hat{u}^2\right)}{\hat{s}^2 N_c^3 C_F}
\begin{bmatrix}
 1 & -C_F \\
 -C_F & C_F^2 \\
\end{bmatrix}\,,
\\
    \bm{H}^{\rm Sivers}_{q\bar{q}\rightarrow \bar{q}q} =&\,\bm{H}^{\rm Sivers}_{q\bar{q}'\rightarrow \bar{q}'q}+\bm{H}^{\rm Sivers}_{q\bar{q}\rightarrow \bar{q}'q'}-\frac{8 \hat{t}^2 g_s^4}{\hat{s} \hat{u} N_c^2 C_F}
\begin{bmatrix}
 2 & -\frac{1}{2} \left(3-N_c^2\right) C_F \\
 -\frac{1}{2} \left(N_c^2+3\right) C_F & C_F^2 \\
\end{bmatrix}\,.
\end{align}
After performing charge conjugation, the hard color matrices for the subprocesses in the bottom row of Tab.\ \ref{tab:quark-channels} can be obtained from these expressions.
\subsubsection{Two quarks and two gluon subprocesses}
All twelve of the two quark and two gluon subprocesses are given in Tab.~\ref{tab:quark-gluon-channels}. As we have mentioned in Sec.~\ref{Introduction}, we neglect the gluon Sivers contribution in this paper. This means that all subprocesses with a gluon incoming from the polarized proton will be neglected. There are then six remaining subprocesses to compute. However, we find that under charge conjugation, the polarized hard functions once again only change by an overall minus sign. Thus, we only need to perform the calculation for three of the hard matrices.

In order to further demonstrate our method for calculating the polarized hard matrices, we now perform the calculation for the $q\bar{q}\rightarrow gg$ subprocess. We then provide the expressions for the remaining hard matrices. For the unpolarized process the scattering amplitude has three channels. After the addition of the external gluon, there are then nine polarized process to be considered. At the cross section level, this results in 27 hard interactions which need to be considered. Despite this complication, we can once again write
\begin{align}
    |\mathcal{M}^{\rm Sivers}|^2 = \frac{1}{C_F}\mathcal{M}^{\textrm{Sivers}\, a}_{j} t^{a}_{j1} \mathcal{M}^{\dagger} \,.
\end{align}
Just like in the unpolarized case, we begin the calculation by decomposing the amplitudes into the primed basis first. Then to simplify our result, we rotate into the unprimed basis. The scattering amplitudes for the process can then be written as
\begin{align}
    \mathcal{M}_{j}^{\textrm{Sivers}\, a}t^{a}_{j1} = \mathcal{M}_{1}^{\rm Sivers}\theta_{1}'
    +\mathcal{M}_{2}^{\rm Sivers}\theta_{2}'+\mathcal{M}_{3}^{\rm Sivers}\theta_{3}'\,,
\end{align}
\begin{align}
    \mathcal{M}_{i}^{\dagger} = \mathcal{M}_1^{\dagger}{\theta_{1}'}^{\dagger}
    +\mathcal{M}_2^{\dagger}{\theta_{2}'}^{\dagger}+\mathcal{M}_3^{\dagger}{\theta_{3}'}^{\dagger}\,,
\end{align}
where
\begin{align}
    \mathcal{M}_1^{\rm Sivers} = \frac{\Tr\left[\mathcal{M}_{j}^{\textrm{Sivers}, a} t^{a}_{j1} \, {\theta_{1}'}^{\dagger}\right]}{\Tr\left[{\theta_{1}'}\,{\theta_{1}'}^{\dagger}\right]}\,,
    \qquad
    \mathcal{M}_2^{\rm Sivers} =\frac{\Tr\left[\mathcal{M}_{j}^{\textrm{Sivers}, a} t^{a}_{j1} \, {\theta_{2}'}^{\dagger}\right]}{\Tr\left[{\theta_{2}'}\,{\theta_{2}'}^{\dagger}\right]}\,,
    \qquad
    \mathcal{M}_3^{\rm Sivers} =\frac{\Tr\left[\mathcal{M}_{j}^{\textrm{Sivers}, a} t^{a}_{j1} \, {\theta_{3}'}^{\dagger}\right]}{\Tr\left[{\theta_{3}'}\,{\theta_{3}'}^{\dagger}\right]}\,,
\end{align}
\begin{align}
    \mathcal{M}_1^{\dagger} = \frac{\Tr\left[\mathcal{M}^{\dagger}\, {\theta_{1}'}\right]}{\Tr\left[{\theta_{1}'}\,{\theta_{1}'}^{\dagger}\right]}\,,
    \qquad
    \mathcal{M}_2^{\dagger} =\frac{\Tr\left[\mathcal{M}^{\dagger} \, {\theta_{2}'}\right]}{\Tr\left[{\theta_{2}'}\,{\theta_{2}'}^{\dagger}\right]}\,.
    \qquad
    \mathcal{M}_3^{\dagger} =\frac{\Tr\left[\mathcal{M}^{\dagger} \, {\theta_{3}'}\right]}{\Tr\left[{\theta_{3}'}\,{\theta_{3}'}^{\dagger}\right]}\,.
\end{align}
The hard matrix in the primed basis can then be computed as 
\begin{align}
    {\bm{H}^{\rm Sivers}_{q\bar{q}\rightarrow gg}}' = \frac{1}{C_F}
    \begin{bmatrix}
    \mathcal{M}_1^{\rm Sivers}\,M_{1}^{\dagger} & \mathcal{M}_1^{\rm Sivers}\,M_{2}^{\dagger} & \mathcal{M}_1^{\rm Sivers}\,M_{3}^{\dagger}\\
    \mathcal{M}_2^{\rm Sivers}\,M_{1}^{\dagger} & \mathcal{M}_2^{\rm Sivers}\,M_{2}^{\dagger} & \mathcal{M}_2^{\rm Sivers}\,M_{3}^{\dagger}\\
    \mathcal{M}_3^{\rm Sivers}\,M_{1}^{\dagger} & \mathcal{M}_3^{\rm Sivers}\,M_{2}^{\dagger} & \mathcal{M}_3^{\rm Sivers}\,M_{3}^{\dagger}\\
    \end{bmatrix}\,.
\end{align}
In order to obtain the hard matrix in the unprimed basis we apply the transformation
\begin{align}
    \bm{H}_{q\bar{q}\rightarrow gg} = \bm{R}^{\dagger}\, \bm{H}_{q\bar{q}\rightarrow gg}'\, \bm{R}
    \qquad
    \bm{R} = 
    \begin{bmatrix}
    1 & 1 & -1\\
    1 & 1 &  1\\
    2 N_c & 0 & 0
    \end{bmatrix}^{-1}\,.
\end{align}
The final result for all of the two quark and two gluon interactions hard matrices are given by
\begin{align}
    \bm{H}^{\rm Sivers}_{q\bar{q}\rightarrow gg}  = & -\frac{4 g_s^4 \left(\hat{s}^2+\hat{u}^2\right)}{\hat{s} \hat{t}^2 \hat{u} N_c C_F}
    \begin{bmatrix}
 2 \hat{s}^2 N_c C_F & 2 \hat{s} \hat{u} N_c C_F & 0 \\
 -\hat{s} \hat{u} \left(N_c^2+1\right) & -\hat{u}^2 \left(N_c^2+1\right) & 0 \\
 \hat{s}^2 N_c & \hat{s} \hat{u} N_c & 0 \\
    \end{bmatrix}\,,
\\
    \bm{H}^{\rm Sivers}_{qg\rightarrow gq}  = & \frac{4 g_s^4 \left(\hat{s}^2+\hat{t}^2\right)}{\hat{s} \hat{t} \hat{u}^2 N_c C_F}
    \begin{bmatrix}
 2 \hat{s}^2 N_c C_F & 2 \hat{s} \hat{t} N_c C_F & 0 \\
 -\hat{s} \hat{t} \left(N_c^2+1\right) & -\hat{t}^2 \left(N_c^2+1\right) & 0 \\
 \hat{s}^2 N_c & \hat{s} \hat{t} N_c & 0 \\
    \end{bmatrix}\,,
 \\
    \bm{H}^{\rm Sivers}_{q\bar{q}\rightarrow gg}  = & \frac{4 g_s^4 \left(\hat{t}^2+\hat{u}^2\right)}{\hat{s}^2 \hat{t} \hat{u} N_c C_F}
    \begin{bmatrix}
 \hat{u}^2 \left(N_c^2+1\right) & \hat{t} \hat{u} \left(N_c^2+1\right) & 0 \\
 \hat{t} \hat{u} \left(N_c^2+1\right) & \hat{t}^2 \left(N_c^2+1\right) & 0 \\
 \hat{s} \hat{u} N_c & \hat{s} \hat{t} N_c & 0 \\
    \end{bmatrix}\,,
\\
\end{align}
After performing charge conjugation, the hard color matrices for the remaining subprocesses can be obtained from these expressions.

\subsubsection{Simplification in the one-dimensional color space}
We note that for processes in which the color space is one dimensional, i.e. single color basis in the decomposition, such as Drell-Yan, SIDIS, and color singlet boson-jet processes, the decomposition of scattering amplitude is trivial. We have
\begin{align}
\mathcal{M} = \mathcal{M}^{\rm kin}\,\theta_1\,,
\end{align}
where $\mathcal{M}^{\rm kin} = \mathcal{M}^{\rm kin}_s+\mathcal{M}^{\rm kin}_t+\mathcal{M}_u^{\rm kin}$ in general receives contribution from different channels as above. The kinematic parts can be trivially extracted by 
\begin{align}
    \mathcal{M}^{\rm kin} = \frac{\Tr\left[\mathcal{M} \theta_1^{\dagger}\right]}{\Tr\left[\theta_1 \theta_1^{\dagger}\right]}\,,
    \qquad
    {\mathcal{M}^{\rm kin}}^{\dagger} = \frac{\Tr\left[\mathcal{M}^{\dagger} \theta_1\right]}{\Tr\left[\theta_1 \theta_1^{\dagger}\right]}\,.
\end{align}
Therefore the unpolarized hard matrices can be constructed simply by\footnote{We keep the boldface notations to be consistent, but $\bm{H}$ and $\bm{S}$ are just numbers here.}
\begin{align}
    \bm{H} = \left| \mathcal{M}^{\rm kin} \right|^2
    \begin{bmatrix}
    1
    \end{bmatrix}\,,
    \qquad
    \bm{S} = 
    \begin{bmatrix}
    \Tr\left[\theta_1 \theta_1^{\dagger}\right]
    \end{bmatrix}\,.
\end{align}
In these expressions, we have suppressed the subprocess subscript since these expressions are true for all subprocesses with a one-dimensional color space. The differential cross section is then given by
\begin{align}
    \left|\mathcal{M} \right|^2 &= \Tr\left[\bm{H} \cdot \bm{S}\right] = C^u \left| \mathcal{M}^{\rm kin} \right|^2
\end{align}
where in the second line we have defined $C^u = \Tr\left[\theta_1 \theta_1^{\dagger}\right]$.
Similarly, for the polarized hard matrix, we can write
\begin{align}
     \left|\mathcal{M}^{\rm Sivers} \right|^2= \frac{\Tr\left[\mathcal{M}^{\rm Sivers, a} t^{a}_{j1} \theta_1^{\dagger}\right]}{\Tr\left[\theta_1 \theta_1^{\dagger}\right]}= \Tr\left[\bm{H}^{\rm Sivers} \cdot \bm{S}\right] = \frac{C^{\rm Sivers}}{C^u}\left|\mathcal{M}^{\rm kin}\right|^2\,,
\end{align}
where $C^{\rm Sivers}\mathcal{M}^{\rm kin} = \Tr\left[\mathcal{M}^{\rm Sivers, a} t^{a}_{j1} \theta_1^{\dagger}\right]$. Therefore, the hard functions of the polarized and unpolarized scatterings are related by an overall color constant,
\begin{align}
    \bm{H}^{\rm Sivers} = \frac{C^{\rm Sivers}}{C^u} \bm{H}\,.
\end{align}
Here, $C^{\rm Sivers}$ can further be decomposed into color factors arising from gauge link gluons interacting with different external colored partons, as seen in \cite{Gamberg:2010tj,Qiu:2007ey,DAlesio:2017rzj,DAlesio:2018rnv}.

\subsection{Evolution equations}
Hard functions can be related to the Wilson coefficients $C_I^\Gamma$ in the color basis $\{\theta_I\}$ of section\ \ref{Hard Functions} by $H_{IJ} = \sum_\Gamma C_I^\Gamma C_J^{\Gamma *}$. Here $\Gamma$ represents different helicity states of the incoming and outgoing particles. Explicit expressions of the Wilson coefficients at next-to-leading order can be found in \cite{Kelley:2010fn,Liu:2014oog}, but we do not present them as we are only using the tree-level hard functions 
for our study. We do, however, include the renormalization group (RG) evolution of the hard functions coming from the $1$-loop anomalous dimensions. Then the Wilson coefficients satisfy the RG evolution equations \cite{Kelley:2010fn,Liu:2014oog,Becher:2009qa,Sterman:2002qn}
\bea
\label{eq:WCRG}
\mu\frac{d}{d\mu}C_I^\Gamma = \left[\left(\gamma_{\rm cusp}\frac{c_H}{2}\ln\frac{-\hat t}{\mu^2} +\gamma_H \right)\delta_{IJ} + \gamma_{\rm cusp} M_{IJ}\right] C_J^\Gamma\,.
\eea
Here, $\gamma_{\rm cusp} = \frac{\alpha_s}{\pi} + \cdots$ is the cusp anomalous dimensions and $c_H = C_a + C_b + C_c + C_d$. The non-cusp anomalous dimension is defined as \bea
\gamma_H=-\frac{1}{2}\left( \gamma_\mu^a\left[\alpha_s(\mu)\right]+\gamma_\mu^b\left[\alpha_s(\mu)\right]+\gamma_\mu^c\left[\alpha_s(\mu)\right]+\gamma_\mu^d\left[\alpha_s(\mu)\right]\right)\,,
\eea
where $\gamma_\mu^i[\alpha_s(\mu)] = \frac{\alpha_s}{\pi} \gamma_i +\cdots$, with $\gamma_q = \frac{3}{2}C_F$ and $\gamma_g = \frac{\beta_0}{2}$. Lastly, the matrix $\bm{M}$ takes the form
\bea
\label{eq:Mmatrix}
\bm{M} = - \sum_{i < j} \bm{T}_i\cdot\bm{T}_j\left[L(s_{ij}) - L(\hat t)\right]\,,
\eea
where $s_{12}=s_{34} = \hat{s}$, $s_{13}=s_{24}=\hat{t}$, and $s_{14} = s_{23} = \hat{u}$ and 
\bea
L(\hat{t}) = \ln \left(\frac{-\hat{t}}{\mu^2}\right)\,,
\qquad
L(\hat{u}) = \ln \left( \frac{-\hat{u}}{\mu^2}\right)\,,
\qquad
L(\hat{s}) = \ln\left( \frac{\hat{s}}{\mu^2}\right) - i\pi\,.
\eea
From the RG evolution of the Wilson coefficients given in Eq.\ \eqref{eq:WCRG}, we can arrive at the RG evolution equations for hard matrix $\bm{H}$ as
\bea
\label{eq:HmuRG}
\mu \frac{d}{d\mu}\bm{H} = \bm{\Gamma}^H\cdot\bm{H} + \bm{H} \cdot\bm{\Gamma}^{H\dagger}\,,
\eea
where $\bm{\Gamma}^H$ is given by
\bea
\label{eq:gammaHmu}
\bm{\Gamma}^H =  \left(\gamma_{\rm cusp}\frac{c_H}{2}\ln\frac{-\hat{t}}{\mu^2} +\gamma_H\right)\bm{I}  + \gamma_{\rm cusp} \bm{M}\,.
\eea

\section{QCD resummation and evolution formalism}
\label{sec:resummation}
In this section, we present the renormalization group (RG) equations for the rest of the key ingredients in the factorized formalism. These include the TMD PDFs, global soft functions, jet functions, and collinear-soft functions. After presenting their NLO perturbative results and RG evolution equations, we check the RG consistency. In the end, we present our resummation formula for dijet production.  

\subsection{TMDs and global soft functions}\label{sec:softfunction}
The unsubtracted TMD PDFs in the factorized formula in Eq.\ \eqref{eq:bunpol} describe the radiation along the incoming beams. They satisfy the RG evolution equations 
\bea
\mu\frac{d}{d\mu} \ln f_i^{\rm unsub}(x,b,\mu,\nu) &= \gamma_\mu^{f_i}(\mu,\nu)\,,\\
\nu\frac{d}{d\nu}\ln f_i^{\rm unsub}(x,b,\mu,\nu) &=\gamma_\nu^{f_i}(\mu)\,,
\eea
where its $\mu$- and $\nu$-anomalous dimensions are given by
\bea
\label{eq:gammafmu}
\gamma_\mu^{f_i}(\mu,\nu)&=\gamma_{\rm cusp}C_i\,\ln \frac{\nu^2}{x_i^2P^{-2}}+ \gamma_\mu^i[\alpha_s(\mu)]\,,\\
\gamma_\nu^{f_i}(\mu,\nu)&=\frac{\alpha_s C_i}{\pi}\ln\frac{\mu^2}{\mu_b^2}\,.
\label{eq:gammafnu}
\eea
As we will see in this subsection, the rapidity divergences of the unsubtracted TMDs will be exactly canceled by the rapidity divergences of the global soft functions, which will allow us to identify the standard TMDs with subtracted rapidity divergence as in Eq.\ \eqref{eq:proper} above. 

Suppressing the label $ab\rightarrow cd$ for convenience, the global soft functions up to $1$-loop are given by
\bea
\label{eq:gsoft}
\bm{S}^{(0) }(b) &= \bm{I}\,,\\
\bm{S}^{\rm{bare},(1) }(b) &= \sum_{i<j}\bm{T}_i\cdot\bm{T}_j\,\mathcal{I}_{ij}^{(1)}(b)\,,
\label{eq:gsoft1loop}
\eea
where \cite{Hornig:2017pud}
\bea
\mathcal{I}_{12}^{(1)}(b) &= \frac{\alpha_s}{2\pi}\left[2\left(\frac{2}{\eta}+\ln\frac{\nu^2}{\mu^2}\right)\left(\frac{1}{\epsilon}+\ln\frac{\mu^2}{\mu_b^2}\right) -\frac{2}{\epsilon^2} + \ln^2\frac{\mu^2}{\mu_b^2}+\frac{\pi^2}{6}\right]\,,\\
\mathcal{I}_{13}^{(1)}(b) &= \frac{\alpha_s}{2\pi}\left[\left(\frac{2}{\eta}+\ln\frac{\nu^2}{\mu^2}-2y_c\right)\left(\frac{1}{\epsilon}+\ln\frac{\mu^2}{\mu_b^2}\right) -\frac{2}{\epsilon^2} -\frac{1}{\epsilon}\ln\frac{\mu^2}{\mu_b^2}+\frac{\pi^2}{6}\right]\,,\\
\mathcal{I}_{34}^{(1)}(b) &= \frac{\alpha_s}{2\pi}\left[4\left(\frac{1}{\epsilon}+\ln\frac{\mu^2}{\mu_b^2}\right)\ln\big(2\cosh(\Delta y/2)\big) -\frac{2}{\epsilon^2}-\frac{2}{\epsilon}\ln\frac{\mu^2}{\mu_b^2} -\ln^2\frac{\mu^2}{\mu_b^2} + \Delta y^2 - 4\ln^2\big(2\cosh(\Delta y/2)\big)+\frac{\pi^2}{6}\right]\,,\\
&\mathcal{I}_{14}^{(1)}(b) = \mathcal{I}_{13}^{(1)}(b)(y_c \to y_d)\,, \qquad  \mathcal{I}_{23}^{(1)}(b) = \mathcal{I}_{13}^{(1)}(b)(y_c \to -y_c)\,, \qquad \mathcal{I}_{24}^{(1)}(b) = \mathcal{I}_{14}^{(1)}(b)(y_d \to -y_d)\,.
\label{eq:gsoftIexp}
\eea
The explicit matrix forms of tree-level soft functions in Eq.\ \eqref{eq:gsoft} for some color basis $\{\theta_I\}$ can be computed as
\bea
(I)_{IJ} = \theta_I \theta_J^\dagger\,,
\eea
which is equivalent to the matrix forms of the LO soft functions found in section\ \ref{Hard Functions}.
The matrix $\bm{T}_i\cdot\bm{T}_j$ of the eq.~\eqref{eq:gsoft1loop} was also computed in the color bases used in section~\ref{Hard Functions} and can be found in \cite{Liu:2014oog,Kelley:2010fn}. The renormalized global soft functions satisfy the RG evolution equations
\bea
\label{eq:globalmuRG}
\mu\frac{d}{d\mu}\bm{S}(b,\mu,\nu) &=\bm{\Gamma}_\mu^{S\dagger}\cdot\bm{S} + \bm{S}\cdot\bm{\Gamma}^S_{\mu}\,,\\
\nu\frac{d}{d\nu}\bm{S}(b,\mu,\nu) &=\bm{\Gamma}_\nu^{S\dagger}\cdot\bm{S} + \bm{S}\cdot\bm{\Gamma}^S_{\nu}\,,\\
\eea
From Eqs.\ \eqref{eq:gsoft} - \eqref{eq:gsoftIexp} and using $\sum_i \bm{T}_i = 0$, we then find
\bea
\label{eq:gammaglobalmu}
\bm{\Gamma}^S_{\mu} =& - \frac{\alpha_s}{2\pi}\left[C_a\left(\ln\frac{-\hat{t}}{x_a^2S} + \ln\frac{\nu^2}{\mu^2}\right) + C_b\left(\ln\frac{-\hat{t}}{x_b^2S} + \ln\frac{\nu^2}{\mu^2}\right) + (C_c+C_d)\left(\ln\frac{-\hat{t}}{P_\perp^2}-\ln\frac{\mu^2}{\mu_b^2}\right)\right]\bm{I}\notag\\
&-\frac{\alpha_s}{\pi}\bm{M} + \frac{\alpha_s}{\pi}\left(\bm{T}_1\cdot\bm{T}_2+\bm{T}_3\cdot\bm{T}_4\right)i\pi\,\notag\\
=& - \frac{\gamma_{\rm cusp}}{2}\left[C_a\left(\ln\frac{-\hat{t}}{x_a^2S} + \ln\frac{\nu^2}{\mu^2}\right) + C_b\left(\ln\frac{-\hat{t}}{x_b^2S} + \ln\frac{\nu^2}{\mu^2}\right) + (C_c+C_d)\left(\ln\frac{-\hat{t}}{P_\perp^2}-\ln\frac{\mu^2}{\mu_b^2}\right)\right]\bm{I}\notag\\
&-\gamma_{\rm cusp}\bm{M}+ \gamma_{\rm cusp}\left(\bm{T}_1\cdot\bm{T}_2+\bm{T}_3\cdot\bm{T}_4\right)i\pi\,,\\
\bm{\Gamma}^S_{\nu} =& -\frac{\alpha_s(C_a+C_b)}{2\pi}\ln\frac{\mu^2}{\mu_b^2}\bm{I}\,,
\eea
where $\bm{M}$ was given in Eq.~\eqref{eq:Mmatrix} and we promoted $\frac{\alpha_s}{\pi} \to \gamma_{\rm cusp}$, which is consistent with the factorization consistency relation below. Note that Eq.~\eqref{eq:gammaglobalmu} is strictly real and the imaginary term $\sim i\pi$ cancels exactly with the imaginary term found in $\bm{M}$.

We note that $\bm{\Gamma}^S_{\nu} \sim \bm{I}$ and that this is expected as the hard functions do not have any rapidity divergence. Thus, we can write
\bea
\label{eq:globalnu}
\nu\frac{d}{d\nu}\bm{S}(b,\mu,\nu) &=\bm{\Gamma}_\nu^{S\dagger}\cdot\bm{S} + \bm{S}\cdot\bm{\Gamma}^S_{\nu} =- \frac{\alpha_s(C_a+C_b)}{\pi}\ln\frac{\mu^2}{\mu_b^2} \bm{S}(b,\mu,\nu)\,,
\eea
which has the same rapidity anomalous dimensions as the back-to-back soft functions $S_{ab}(b,\mu,\nu)$ found in standard Drell-Yan and SIDIS process~\cite{Chiu:2012ir}. As expected, the rapidity divergence of the global soft function $\bm{S}(b,\mu,\nu)$ in Eq.~\eqref{eq:globalnu} exactly cancels the rapidity anomalous dimensions for the unsubtracted TMDs $f_a(b,\mu,\nu)$ and $f_b(b,\mu,\nu)$ given in Eq.~\eqref{eq:gammafnu}. Therefore, as discussed in the introduction, we can define $\tilde{\bm{S}}(b,\mu)$ absent of the rapidity divergence such that
\bea
\bm{S}(b,\mu,\nu) = \tilde{\bm{S}}(b,\mu)S_{ab}(b,\mu,\nu)\,.
\eea
Then as in Eq.\ \eqref{eq:proper}, $S_{ab}(b,\mu,\nu)$ is combined with the unsubtracted TMDs to identify standard TMDs free of the rapidity divergences. 

\subsection{Jet and collinear-soft functions}
Both jet and collinear-soft functions describe the radiation which resolves the produced jets. The jet functions \cite{Ellis:2010rwa,Liu:2012sz} encode the collinear radiations inside anti-$k_T$ jet with radius $R$. The NLO expressions are given by
\bea
J_i(P_\perp R,\mu) = 1+\frac{\alpha_s}{\pi}\left[\frac{C_i}{4} \ln^2\left(\frac{\mu^2}{P_\perp^2 R^2}\right) +\frac{\gamma_i}{2}\ln\left(\frac{\mu^2}{P_\perp^2R^2}\right)+d_i\right]\,,
\eea
where the algorithmic dependent terms $d_i$ for anti-$k_T$ algorithm are
\bea
d_q &= \left(\frac{13}{4}-\frac{3\pi^2}{8}\right)C_F\,,\\
d_g &= \left(\frac{67}{18}-\frac{3\pi^2}{8}\right)C_A - \frac{23}{36}n_f\,.
\eea
The jet functions satisfy the RG evolution equations
\bea
\mu\frac{d}{d\mu} J_i(P_\perp R,\mu) = \gamma_\mu^{J_i}(\mu) J_i(P_\perp R,\mu)\,,
\eea
where the anomalous dimension is given by
\bea
\label{eq:gammajetmu}
\gamma_\mu^{J_i}(\mu) = \gamma_{\rm cusp}C_i\,\ln \left(\frac{\mu^2}{P_\perp^2R^2}\right)+ \gamma_\mu^i[\alpha_s(\mu)]\,.
\eea

The collinear-soft functions \cite{Buffing:2018ggv,Chien:2019gyf} describe the soft radiation along the jet direction and resolves the jet cone $R$. The NLO expressions are given by
\bea
S^{cs,(1)}_i(b,R,\mu) = 1-\frac{\alpha_s C_i}{4\pi}\left[ \ln^2\left(\frac{\mu^2}{\mu_b^2 R^2}\right) -\frac{\pi^2}{6}\right]\,.
\eea
The collinear-soft functions satisfy the RG evolution equations
\bea
\mu\frac{d}{d\mu} S^{\rm cs}_i(b,R,\mu) = \gamma_\mu^{cs_i}(\mu) S^{\rm cs}_i(b,R,\mu) \,,
\eea
where its anomalous dimension takes the form
\bea
\label{eq:gammacsmu}
\gamma_\mu^{cs_i}(\mu)=\gamma_{\rm cusp}C_i\,\ln \left(\frac{\mu^2}{\mu_b^2R^2}\right)\,.
\eea

\subsection{RG consistency at $1$-loop}\label{sec:consistency}
With the anomalous dimensions presented for all the ingredients, we now show that our factorized formula given in Eq.\ \eqref{eq:bunpol} satisfy the consistency relations for the RG evolutions. The cancellation of the rapidity divergences was already checked around Eq.\ \eqref{eq:globalnu}. We also expect $\mu$-divergence of the various functions to cancel and satisfy the consistency equation
\bea
\label{eq:muconsistency}
\mu\frac{d}{d\mu}\ln\big({\rm Tr}\left[ \bm{S}(b,\mu,\nu)\cdot
\bm{H}(P_{\perp}, \mu) \right] \big) + \gamma^{f_a}_\mu+ \gamma^{f_b}_\mu+ \gamma^{cs_c}_\mu+ \gamma^{cs_d}_\mu + \gamma^{J_c}_\mu +\gamma^{J_d}_\mu &=0 \,.
\eea

From Eqs.\ \eqref{eq:HmuRG}, \eqref{eq:gammaHmu}, \eqref{eq:globalmuRG}, \eqref{eq:gammaglobalmu}, we immediately find at $1$-loop,
\bea
\mu\frac{d}{d\mu}\ln\big({\rm Tr}\left[ \bm{S}(b,\mu,\nu)\cdot
\bm{H}(P_{\perp}, \mu) \right] \big) &=\frac{{\rm Tr}\left[\bm{\Gamma}_\mu^{S\dagger}\cdot\bm{S}\cdot\bm{H} + \bm{S}\cdot\bm{\Gamma}^S_{\mu}\cdot\bm{H} + \bm{S}\cdot\bm{\Gamma}^H\cdot\bm{H} +\bm{S}\cdot\bm{H}\cdot\bm{\Gamma}^{H\dagger}\right]}{{\rm Tr}\left[ \bm{S}(b,\mu,\nu)\cdot
\bm{H}(P_{\perp}, \mu) \right] }\notag\\
&= -\frac{\alpha_s}{\pi}\left[C_a\ln\left(\frac{\nu^2}{x_a^2S}\right) + C_b\ln\left(\frac{\nu^2}{x_b^2S}\right) - (C_c+C_d)\ln\left(\frac{P_\perp^2}{\mu_b^2}\right)\right]+2\gamma_H\,.
\eea
One can then easily check from the $\mu$-anomalous dimensions of the other functions given in Eqs.\ \eqref{eq:gammafmu}, \eqref{eq:gammajetmu}, \eqref{eq:gammacsmu} that Eq.\ \eqref{eq:muconsistency} is explicitly satisfied at $1$-loop.

\subsection{Resummation formula}\label{resum}
Based on the above discussions and RG renormalization group methods in SCET, we can now derive the expression for the all-order resummed result. Explicitly, we calculate the cross section at the NLL accuracy, where we will use the two-loop cusp and one-loop single logarithmic anomalous dimension and the matching coefficients are kept at leading order. On the other hand, the color structures inside the hard and soft function will mix with each other under the RG evolution, which was first studied in \cite{Kidonakis:1998nf}. In this paper, we will apply the same methods in~\cite{Kelley:2010fn} to solve the RG equations. For the unpolarized cross section, the resummation formula has the form as follows:
\begin{align}
\frac{d\sigma}{dy_c dy_d d P_\perp^2 d^2\vec q_\perp } =& \sum_{abcd} \frac{1}{16\pi^2 \hat{s}^2} \frac{1}{N_{\rm init}} \frac{1}{1+\delta_{cd}} \frac{1}{2\pi}\int_0^{\infty} db \,b\,J_0(q_\perp b) \,x_a f_a(x_a,\mu_{b_{*}})\, x_b f_b(x_b,\mu_{b_{*}}) \notag \\
&\times  
\exp\left\{ - \int_{\mu_{b_{*}}}^{\mu_h} \frac{d\mu}{\mu} \left[ \gamma_{\rm cusp}(\alpha_s)c_H \,\ln \frac{|\hat t|}{\mu^2}+ 2\gamma_H(\alpha_s) \right] \right\} \notag \\
&\times \sum_{KK'} \exp\left[ -\int_{\mu_{b_{*}}}^{\mu_h} \frac{d\mu}{\mu} \gamma_{\rm cusp}(\alpha_s)(\lambda_K+\lambda_{K'}^*) \right] H_{KK'}(P_{\perp}, \mu_h)  \tilde{S}_{K'K}(b_{*},\mu_{b_{*}})  \notag \\  
&\times \exp\left[ - \int_{\mu_{b_{*}}}^{\mu_j} \frac{d\mu}{\mu} \gamma_\mu^{J_c}(\alpha_s) - \int_{\mu_{b_{*}}}^{\mu_{cs}} \frac{d\mu}{\mu} \gamma_\mu^{cs_c}(\alpha_s) \right] U_{\mathrm{NG}}^{c}\left(\mu_{cs}, \mu_{j}\right) J_c(P_\perp R,\mu_j)  S^{\rm cs}_c(b_{*},R,\mu_{cs})\notag \\
&\times \exp\left[ - \int_{\mu_{b_{*}}}^{\mu_j} \frac{d\mu}{\mu} \gamma_\mu^{J_d}(\alpha_s) - \int_{\mu_{b_{*}}}^{\mu_{cs}} \frac{d\mu}{\mu} \gamma_\mu^{cs_d}(\alpha_s) \right]  U_{\mathrm{NG}}^{d}\left(\mu_{cs}, \mu_{j}\right) J_d(P_\perp R,\mu_j)  S^{\rm cs}_d(b_{*},R,\mu_{cs})\,, \notag \\
& \times \exp\left[-S^{a}_{\rm NP}(b,Q_0,\sqrt{\hat s}) -S^{b}_{\rm NP}(b,Q_0,\sqrt{\hat s}) \right],
\label{eq:unpol-resum}
\end{align}
where $\lambda_K$ is the eigenvalue of the matrix $M_{IJ}$ in the hard anomalous dimension \eqref{eq:WCRG} and $H_{KK'}$ and $\tilde{S}_{K'K}$ are the hard and soft function in the diagonal basis as defined in \cite{Kelley:2010fn}. In our numerical calculation, we use the LAPACK library \cite{lapack99} to obtain their value at different phase-space points. We have applied the $b_*$-prescription to prevent the Landau pole from being reached in the $b$-integral. Here, we define $b_{*}$ as
\begin{align}
b_{*}=b/\sqrt{1+b^2/b_{\rm max}^2}\,,
\end{align}
where $b_{\rm max}$ is chosen~\cite{Kang:2015msa} to be 1.5~GeV$^{-1}$. The nonperturbative Sudakov factor in Eq.~\eqref{eq:unpol-resum} was fitted to experimental data in \cite{Su:2014wpa}. The extracted functions are given by
\begin{align}
    S_{\rm NP}^{a,b}(b,Q_0,\mu) = g_1^f b^2+\frac{g_2}{2} \frac{C_{a,b}}{C_F}\ln{\frac{\mu}{Q_0}}\ln{\frac{b}{b_*}},\quad {\rm with}~ g_1^f = 0.106,~ g_2=0.84,~ Q_0^2=2.4\,{\rm GeV}^2.
\end{align}
We also incorporate NGLs resummation effects included by the function $U_{\mathrm{NG}}^{c,d}$. In order to include NGLs resummation effects at NLL accuracy, we also need to consider the extra one-loop single logarithmic anomalous dimension $\bm{\hat \Gamma}$ from the non-linear evolution parts. However, in \cite{Becher:2015hka,Becher:2016mmh} this anomalous dimension was shown to cancel between the jet and collinear-soft function up to two-loop order. The explicit operator-based derivation of RG consistency including $\bm{\hat \Gamma}$ can be found in \cite{Becher:2016omr, Chien:2019gyf,Kang:2020yqw}.  In the large $N_c$ limit, the non-linear evolution equation can be solved using the parton shower algorithm \cite{Balsiger:2018ezi}.  Especially, at the NLL accuracy the evolution is totally determined by the one-loop anomalous dimension $\bm{\hat \Gamma}$, which is equivalent to the one appearing in the light jet mass distribution at the $e^+ e^-$ collider. Therefore, we can use the same fitting function form given in \cite{Dasgupta:2001sh} to capture NGLs resummation contributions after setting proper initial and final evolution scales. In our case, these two scales are the jet scale $\mu_j$ and the collinear-soft scale $\mu_{cs}$. Explicitly, the function is
\begin{align}
    U_{\mathrm{NG}}^{k}\left(\mu_{cs}, \mu_{j}\right) = \exp \left[-C_{A} C_{k} \frac{\pi^{2}}{3} u^{2} \frac{1+(a u)^{2}}{1+(b u)^{c}}\right],
\end{align}
where the superscript $k=q$ and $g$ denote the (anti-)quark and gluon jet, respectively, and with $C_q=C_F$ and $C_g=C_A$. The parameters $a$, $b$ and $c$ are fitting parameters which are given as $a=0.85\,C_A$, $b=0.86\,C_A$ and $c=1.33$. The variable $u=\frac{1}{\beta_{0}} \log \frac{\alpha_{s}\left(\mu_{cs}\right)}{\alpha_{s}\left(\mu_{j}\right)}$ is the evolution scale measuring the separation of the scales $\mu_{cs}$ and $\mu_j$.

As we have done for the unpolarized cross section, we also derive a similar resummation formula for the spin-dependent cross section 
\begin{align}
    \frac{d\Delta\sigma(S_\perp)}{dy_c dy_d d P_\perp^2 d^2\vec q_\perp } =\,& \sin(\phi_q - \phi_S)\sum_{abcd} \frac{1}{16\pi^2 \hat{s}^2} \frac{1}{N_{\rm init}} \frac{1}{1+\delta_{cd}} \left(-\frac{1}{4\pi}\right) \int_0^\infty db \, b^2\, J_1(q_\perp b)\, x_a T_{a,F}(x_a, x_a ,\mu_{b_{*}})\, x_b f_b(x_b,\mu_{b_{*}}) \notag \\
    &\times  
\exp\left\{ - \int_{\mu_{b_{*}}}^{\mu_h} \frac{d\mu}{\mu} \left[ \gamma_{\rm cusp}(\alpha_s)c_H \,\ln \frac{|\hat t|}{\mu^2}+ 2\gamma_H(\alpha_s) \right] \right\} \notag \\
&\times \sum_{KK'} \exp\left[ -\int_{\mu_{b_{*}}}^{\mu_h} \frac{d\mu}{\mu} \gamma_{\rm cusp}(\alpha_s)(\lambda_K+\lambda_{K'}^*) \right] H_{KK'}(P_{\perp}, \mu_h)  \tilde{S}_{K'K}(b_{*},\mu_{b_{*}})  \notag \\  
&\times \exp\left[ - \int_{\mu_{b_{*}}}^{\mu_j} \frac{d\mu}{\mu} \gamma_\mu^{J_c}(\alpha_s) - \int_{\mu_{b_{*}}}^{\mu_{cs}} \frac{d\mu}{\mu} \gamma_\mu^{cs_c}(\alpha_s) \right] U_{\mathrm{NG}}^{c}\left(\mu_{cs}, \mu_{j}\right) J_c(P_\perp R,\mu_j)  S^{\rm cs}_c(b_{*},R,\mu_{cs})\notag \\
&\times \exp\left[ - \int_{\mu_{b_{*}}}^{\mu_j} \frac{d\mu}{\mu} \gamma_\mu^{J_d}(\alpha_s) - \int_{\mu_{b_{*}}}^{\mu_{cs}} \frac{d\mu}{\mu} \gamma_\mu^{cs_d}(\alpha_s) \right]  U_{\mathrm{NG}}^{d}\left(\mu_{cs}, \mu_{j}\right) J_d(P_\perp R,\mu_j)  S^{\rm cs}_d(b_{*},R,\mu_{cs})\,, \notag \\
& \times \exp\left[-S^{s}_{\rm NP}(b,Q_0,\sqrt{\hat s}) -S^{b}_{\rm NP}(b,Q_0,\sqrt{\hat s}) \right],
\label{eq:pol-resum}
\end{align}
where at the NLL accuracy we keep the LO matching coefficient in Eq.\ (\ref{eq:siver-matching}). It involves the parametrization for the Sivers function, which depends on the collinear Qiu-Sterman function $T_{q,F}(x_a, x_a, \mu_{b_{*}})$ and a different non-perturbative Sudakov factor $S_{\rm NP}^{s}$. The relevant parametrization has been determined from a recent global analysis of the Sivers asymmetry of SIDIS and Drell-Yan processes~\cite{Echevarria:2020hpy}. The non-perturbative Sudakov factor is given by
 \begin{align}
    S_{\rm NP}^{s}(b,Q_0,\mu) = g_1^s b^2+\frac{g_2}{2}\ln{\frac{\mu}{Q_0}}\ln{\frac{b}{b_*}}\,,\quad {\rm with}~g_1^s = 0.18.
\end{align}

\section{Phenomenology}\label{Phenomenology}

In this section we will present the numerical results using the resummation formula in Eqs.\ \eqref{eq:unpol-resum} and \eqref{eq:pol-resum}, where intrinsic scales for the hard, jet and collinear-soft function are chosen as
\begin{align}
    \mu_h = \sqrt{\hat s},\quad \mu_j=P_\perp R,\quad\mu_{cs} = \mu_{b^*} R.
\end{align}
In the numerical study, we will focus on the Sivers asymmetry for the dijet production at the RHIC with $\sqrt{s}=200~{\rm GeV}$, where the jet events are reconstructed by using anti-$k_T$ algorithm with jet radius $R=0.6$. The transverse momentum $P_\perp$ and the rapidity $y_{c,d}$ of jets are   
\begin{align}
    P_\perp >4\,{\rm GeV},~~~-1<y_{c,d}<2.
\end{align}
For the unpolarized proton, we use the HERAPDF20NLO parton distribution functions~\cite{Abramowicz:2015mha}. The numerical Bessel transforms in Eqs.~\eqref{eq:unpol-resum} and \eqref{eq:pol-resum} are performed using the algorithm in \cite{Kang:2019ctl}. Furthermore, the Eq.~\eqref{eq:unpol-final} is derived after neglecting the power corrections from $\mathcal{O}(q_\perp^2/P_\perp^2)$. In other words, in the large $q_\perp$ region, the full results should include corrections from the so-called $Y$-term, which can be obtained from perturbative QCD calculations \cite{Currie:2017eqf}. In this paper we focus on the contribution from back-to-back dijet production. In order to select such kinematics, we require the transverse momentum $q_\perp$ for the dijet system $|q_\perp|< q_\perp^{\rm cut}$. In the numerical calculations, we fix the value of $q_\perp^{\rm cut}=2$ GeV. 

\begin{figure}[t]
    \centering
    \includegraphics[width=0.45\linewidth]{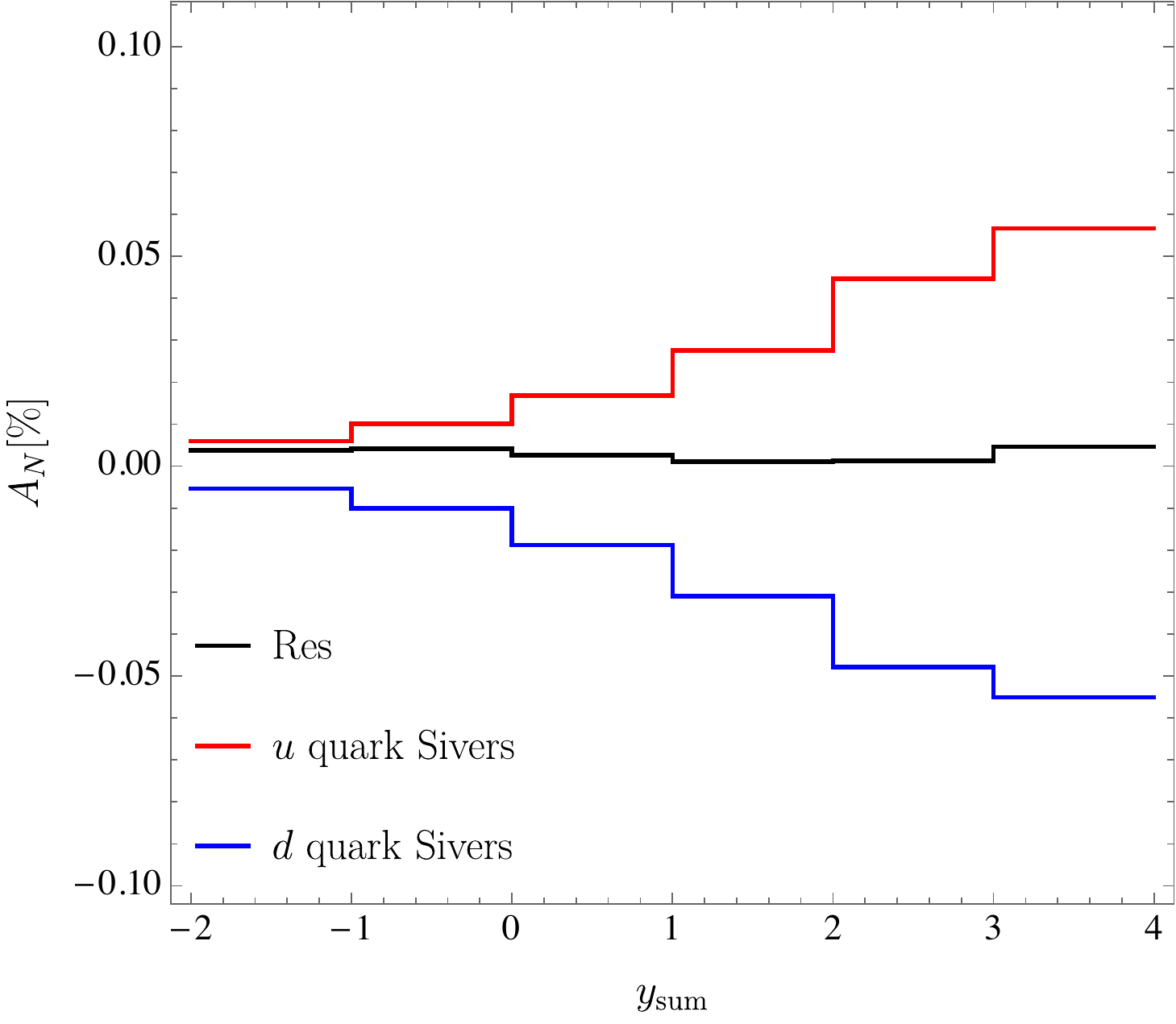} 
    \hspace{1cm}
    \includegraphics[width=0.45\linewidth]{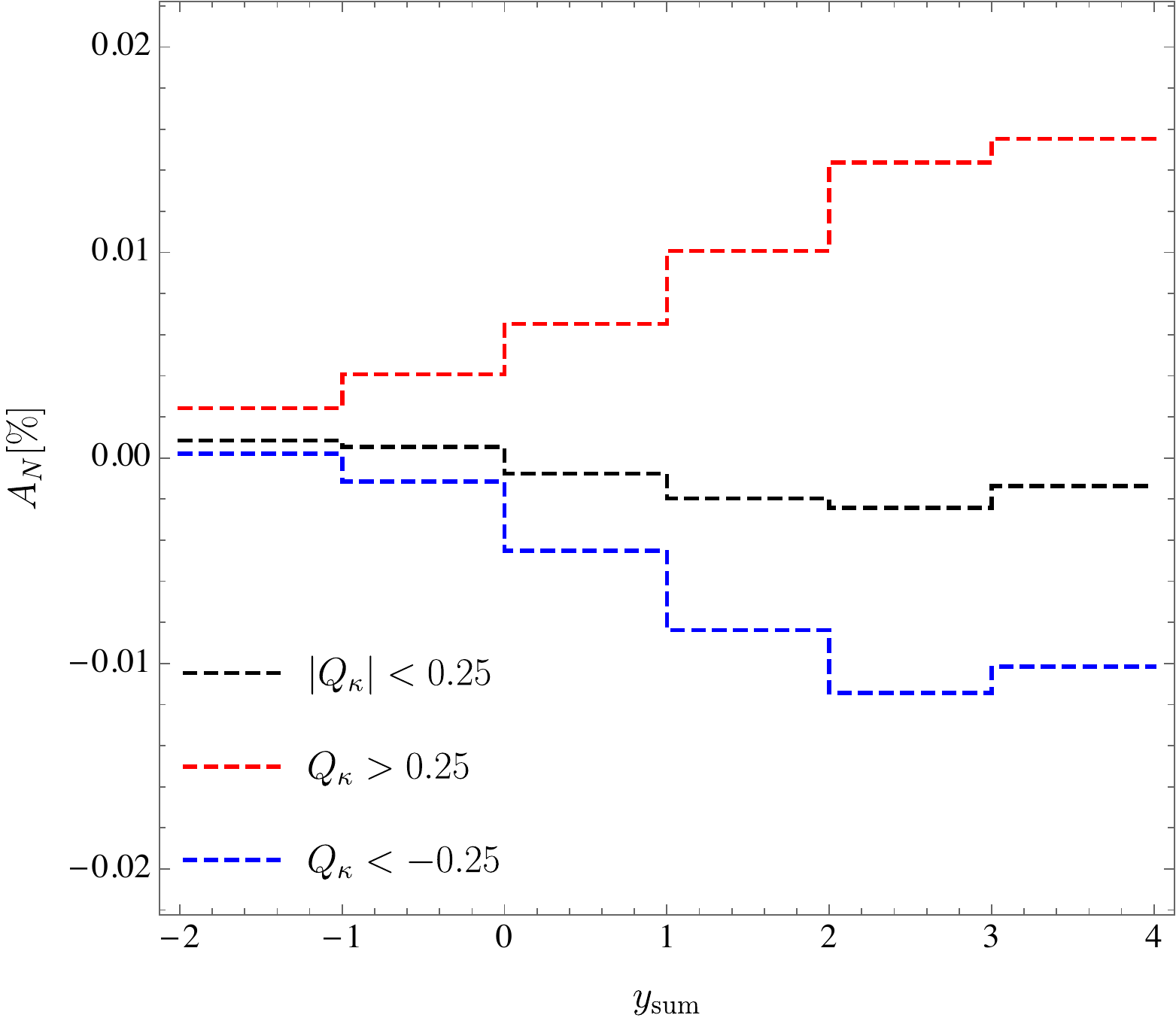} 
    \caption{Theoretical predictions of the Sivers asymmetry for dijet production at the RHIC with $\sqrt{s}=200$ GeV. In the left plot red and blue curves are the results from $u$- and $d$- quark Sivers function, and the black curve includes all the contributions. In the right plot we show the Sivers asymmetry distribution within three different jet charge $Q_\kappa$ bins.}
    \label{fig:SSA}
\end{figure}

As shown in the Fig.~\ref{fig:dijet_fig}, the transverse-polarized proton moves on $+z$-direction and its spin points to $+y$-direction with $\phi_S=\pi/2$. The transverse momentum vector $\vec q_\perp$ lies in the $x-y$ plane, and the Sivers asymmetry is defined as the difference of the events between $q_{\perp,x}>0$ and $q_{\perp,x}<0$ hemispheres, that is the same as the measurements by STAR collaboration \cite{Abelev:2007ii}. Explicitly, we have
\begin{align}\label{eq:AN}
A_N(y_{\rm sum}) = \frac{ \int_0^{q_\perp^{\rm cut}} dq_\perp \int_{0}^{2\pi}d\phi_q \int d\mathcal{PS}  \frac{d\Delta\sigma}{d q_\perp d\phi_q dy_c dy_d dP_\perp} \Big[\theta(\cos\phi_q)  -  \theta(-\cos\phi_q)  \Big]}{ \int_0^{q_\perp^{\rm cut}} d q_\perp \int_{0}^{2\pi}d\phi_q \int d \mathcal{PS} \frac{d\sigma}{d q_\perp d\phi_q dy_c dy_d dP_\perp}    },
\end{align}
with $\int d \mathcal{PS} = \int d y_c d y_d d P_\perp \delta(y_{\rm sum}-y_c-y_d)$ represents the transverse momenta and rapidities integral for dijets. In the numerator, the $\phi_q$-integral with $\theta(\cos\phi_q)$ and $\theta(-\cos\phi_q)$ corresponds $q_{\perp,x}>0$ and $q_{\perp,x}<0$, respectively. 

In the Fig.~\ref{fig:SSA}, we show the numerical results of the Sivers asymmetry for dijet processes, where we neglect the charm and bottom jet events. The red and blue curves represent the asymmetry contributed from $u$- and $d$-quark Sivers function, respectively. As is expected, we find that the asymmetry is enhanced in the large $y_{\rm sum}$ region, i.e. the forward scattering region, due to the larger fractional contribution of Sivers function in the valence region. Besides, the contributions from $u$- and $d$-quark Sivers function are opposite from each other, which causes a huge cancellation of the asymmetry, as shown by the black curves in Fig.~\ref{fig:SSA}.    

In the calculation, most of the asymmetries come from the partonic scattering process $qg\to qg$ where the initial quark comes from the polarized proton. Especially, the more forward jet is associated with the parton from the polarized proton moving in the same direction. Hence, if we can tag parton species initiating the more forward jet, then we can separate $u$- and $d$-quark Sivers functions and avoid the accidental cancellation as shown in the left plot of Fig.~\ref{fig:SSA}.  

In order to achieve jet flavor separation mentioned above, one possible method is applying the electric charge information of jets, which has been proposed in \cite{Aschenauer:2015eha,bnltalk,Kang:2020fka}. In this paper, we will use the standard jet electric charge definition given in ~\cite{Krohn:2012fg, Waalewijn:2012sv}
\bea\label{eq:jetcharge}
Q_\kappa = \sum_{h\in {\rm jet}} z_h^\kappa Q_h \,,
\eea
where $z_h$ is the transverse momentum ratio between hadrons and the jet. $\kappa$ is an input parameter, which is fixed by $\kappa=0.3$~\cite{Kang:2020fka} in our calculations. As shown in \cite{Kang:2020fka}, after measuring the jet charge information, the theory formula is slightly modified by replacing the jet function $J_i(P_\perp R,\mu)$ in Eq.~\eqref{eq:pol-final} by the charge-tagged jet function $\mathcal{G}_i(Q_\kappa,P_\perp R,\mu)$ as
\begin{align}
   \frac{d\Delta\sigma}{d Q_\kappa d^2 q_\perp} = \int d\mathcal{PS} \, T_{a,F} \otimes f_b \otimes {\rm Tr}[\bm{H}\cdot\bm{S}]\otimes S^{cs}_c \otimes S^{cs}_d \,\big[\mathcal{G}_c \, J_d \, \theta(y_c-y_d) + J_c \, \mathcal{G}_d \, \theta(y_d-y_c) \big],
\end{align}
with the normalization as $\int_{-\infty}^\infty dQ_\kappa \, {\cal G}_{i}(Q_\kappa,P_\perp R, \mu)=J_{i}(P_\perp R, \mu)$ required by the probability conservation. Here we only replace the more forward jet function with the charge-tagged jet function, which corresponds to the insertion of the step function. We define the jet charge bin fraction as
\begin{align}
    r_i^{\rm bin} = \frac{\int_{\rm bin} d Q_\kappa \, \mathcal{G}_i(Q_\kappa,P_\perp R, \mu)}{J_i(P_\perp R, \mu)}.
\end{align}
Then the Sivers asymmetry $A_N$ in different jet charge bins is given as, in terms of jet charge bin fraction
\begin{align}
A_N^{\pm,0} = \frac{\sum_{i=u,d,g,\cdots} r_i^{\pm,0} \Delta \sigma_i}{\sigma},
\end{align}
where we suppress the phase space integral shown in Eq.~\eqref{eq:AN}. The index $i$ denotes the parton species initiating the more forward jet. Here we use the same jet charge bins defined in \cite{Kang:2020fka}, where $+,-$ and $0$ indicate $Q_\kappa>0.25, Q_\kappa<-0.25$ and $|Q_\kappa|<0.25$ bins, separately. Such jet charge bin fraction can be fitted from the unpolarized cross section for back-to-back dijet events at the RHIC. In \cite{bnltalk}, the authors have shown the preliminary results from the measurements as $\kappa=0$. In the theory calculation, one can use Monte-Carlo event generators such as Pythia8 \cite{Sjostrand:2007gs} to estimate these numbers. In the Tab.~\ref{tab:charge-bin-fraction} we give the results of jet charge bin fractions $r^{\pm,0}_i$ for various jet flavors used in our numerical calculations, where the jet charges are defined using all charged hadrons inside the jet.

\begin{table}
\centering
\begin{tabular}{c | c  c c c c c c}
\hline
\hline
  & $u$ & $ {\bar u}$ & $d$ & ${\bar d}$ & $s$ & ${\bar s} $ & $g$  \  \\ 
\hline
 $r_i^+$ & ~$0.61$ & $ 0.16 $ & $0.15$ & $0.51$ & $0.15$ & $ 0.50 $ & $0.37$  \  \\ 
 $r_i^-$ & ~$ 0.10$ & $ 0.54 $ & $0.48$ & $0.14$ & $0.49$ & $0.16 $ & $0.37$ \  \\ 
 $r_i^0$ & ~$0.29$ & $ 0.30 $ & $0.37$ & $0.35$ & $0.36$ & $0.34 $ & $0.26$ \  \\ 
\hline
\end{tabular}
\caption{The jet charge bin fractions $r^{\pm,0}_i$ for various jet flavors from Pythia8 simulation, where the jet charges are defined using all charged-hadrons inside the jet.}
\label{tab:charge-bin-fraction}
\end{table}

In the right plot of Fig.~\ref{fig:SSA} we show the result of $A_N$ within the different jet charge bins. After selecting the charge of the more forward jet $Q_\kappa>0.25$, the contribution from the $u$-quark Sivers function is enhanced compared to the case without the jet charge measurement (the black curve in the left plot). A similar size enhancement from the $d$-quark Sivers function is also observed in $Q_\kappa<-0.25$ charge bin as shown by the blue curve. Besides, we find the Sivers asymmetries from $Q_\kappa>0.25$ bins are positive and $Q_\kappa<-0.25$ bins are negative, which are consistent with the preliminary STAR measurements \cite{bnltalk}. In the forward region, the Sivers asymmetry can achieve $\mathcal{O}(0.01\%)$, and size of our calculation is also around the same order of the data. Taken together, our calculation suggests that the dijet production at the hadron collider is an important process to extract the information about the Siver function and deserves further studies on the theoretical framework about the remarks discussed in \ref{sec:remark}.

\section{Conclusions}\label{Conclusions}
We study the single spin asymmetries of dijet production in the back-to-back region in transversely polarized proton-proton collisions. In the back-to-back region, the dijet transverse momentum imbalance $q_\perp$ is much smaller than the transverse momentum $P_\perp$ of the jets. In this case, the conventional perturbative QCD calculations in the expansion of coupling constant $\alpha_s$ generate large logarithms in the form of $\alpha_s^n\, \ln^m\left(P_\perp^2/q_\perp^2\right)$ with $m\leq 2n-1$, which have to be resummed in order to render the convergence of the perturbative computations. We propose a QCD formalism in terms of transverse momentum dependent (TMD) parton distribution functions for dijet production in both unpolarized and polarized proton-proton collisions. Such a formalism allows us to resum the aforementioned large logarithms, and further takes into account the non-universality or process-dependence of the Sivers functions in the case of the transversely polarized scattering. It is well-known that hadronic dijet production in back-to-back region suffers from TMD factorization breaking effects. Thus, to write down the QCD ``seemingly factorized'' formalism for resumming large logarithms mentioned above, we make a couple of approximations. First of all, we neglect the Glauber mode in the formalism which are known to be the main reason for the TMD factorization breaking. Secondly, we have assumed that the soft gluon radiation that is encoded in the global soft function in our formalism is spin-independent, i.e., they are the same between the unpolarized and polarized scatterings. Since the precise method for dealing with the TMD factorization breaking effects is still not known, we feel that the proposed formalism in this paper is a reasonable starting point for further investigation.  

With such a formalism at hand, we compute the Sivers asymmetry for the dijet production in the kinematic region that is relevant to the proton-proton collisions at the Relativistic Heavy Ion Collider (RHIC), and find that the spin asymmetry is very small due to the cancellation between $u$- and $d$-quark Sivers functions, which are similar in size but opposite in sign. However, we find that the individual contribution from $u$- and $d$-quark Sivers functions can lead to an asymmetry of size ${\cal O}(\pm 0.05\%)$ in the forward rapidity region, which seems feasible at the RHIC. Motivated by this, we compute the Sivers asymmetry of dijet production in the positive and negative jet charge bins, i.e., when the jet charge $Q_\kappa$ for the jet with the larger rapidity of two is in the bins $Q_\kappa > 0.25$ and $Q_\kappa < -0.25$, respectively. By selecting the positive (negative) jet charge bin, we enhance the contribution from $u$- ($d$)-quark Sivers function and thus enhance the size of the asymmetry. Our calculation shows that Sivers asymmetries in such positive (negative) jet charge bins lead to asymmetries of size ${\cal O}(+0.01\%)$ (${\cal O}(-0.01\%)$), respectively. The sign of such asymmetries seem to be consistent with the preliminary STAR measurements at the RHIC. The size of our calculations is also around the same order of the experimental data. This give us a great hope to further investigate the single spin asymmetries for hadronic dijet production at the RHIC.

\section*{Acknowledgements}
We thank Huanzhao Liu for useful correspondence on the experimental measurements, thank Maarten Buffing for collaborating during the early stages of this project and thank Zelong Liu for useful discussions.  Z.K. and D.Y.S. are supported by the National Sciencez Foundation under Grant No. PHY-1720486 and CAREER award PHY-1945471. K.L is supported by the National Science Foundation under Grant No. PHY-1316617 and No. PHY-1620628. J.T. is supported by NSF Graduate  Research Fellowship Program under Grant No. DGE-1650604. D.Y.S. is also supported by Center for Frontiers in Nuclear Science of Stony Brook University and Brookhaven National Laboratory. This work is supported within the framework of the TMD Topical Collaboration.


\textit{Note added:} While this work was being written up, we noticed a similar work \cite{Liu:2020jjv} appears on arXiv. The authors investigate process dependent factorization violation from the soft gluon radiation. Their method is different from our approach. We assume a factorized form for the spin-dependent cross section, which we demonstrate to be renormalization group consistent. Within this factorized form, we explicitly calculate the process dependent polarized hard function in the matrix form. Besides, in the numerical calculations we include quark Sivers functions in all the partonic channels. We believe these two studies are complementary with each other. 


\bibliography{refs}

\begin{thebibliography}{100}

\bibitem{Boer:2011fh}
D.~Boer {\em et~al.},
\newblock (2011), arXiv:1108.1713.

\bibitem{Accardi:2012qut}
A.~Accardi {\em et~al.},
\newblock Eur. Phys. J. A {\bf 52}, 268 (2016), arXiv:1212.1701.

\bibitem{Aidala:2020mzt}
C.~A. Aidala {\em et~al.},
\newblock {\em {Probing Nucleons and Nuclei in High Energy Collisions}} (WSP,
  2020), arXiv:2002.12333.

\bibitem{Sivers:1989cc}
D.~W. Sivers,
\newblock Phys.\ Rev.\ D {\bf 41}, 83 (1990).

\bibitem{Sivers:1990fh}
D.~W. Sivers,
\newblock Phys. Rev. D {\bf 43}, 261 (1991).

\bibitem{Antille:1980th}
J.~Antille {\em et~al.},
\newblock Phys. Lett. B {\bf 94}, 523 (1980).

\bibitem{Adams:1991rw}
E581, E704, D.~Adams {\em et~al.},
\newblock Phys. Lett. B {\bf 261}, 201 (1991).

\bibitem{Adams:2003fx}
STAR, J.~Adams {\em et~al.},
\newblock Phys. Rev. Lett. {\bf 92}, 171801 (2004), arXiv:hep-ex/0310058.

\bibitem{Arsene:2008aa}
BRAHMS, I.~Arsene {\em et~al.},
\newblock Phys. Rev. Lett. {\bf 101}, 042001 (2008), arXiv:0801.1078.

\bibitem{Abelev:2008af}
STAR, B.~Abelev {\em et~al.},
\newblock Phys. Rev. Lett. {\bf 101}, 222001 (2008), arXiv:0801.2990.

\bibitem{Adamczyk:2012xd}
STAR, L.~Adamczyk {\em et~al.},
\newblock Phys. Rev. D {\bf 86}, 051101 (2012), arXiv:1205.6826.

\bibitem{Adare:2013ekj}
PHENIX, A.~Adare {\em et~al.},
\newblock Phys. Rev. D {\bf 90}, 012006 (2014), arXiv:1312.1995.

\bibitem{Adamczyk:2017wld}
STAR, L.~Adamczyk {\em et~al.},
\newblock Phys. Rev. D {\bf 97}, 032004 (2018), arXiv:1708.07080.

\bibitem{Kane:1978nd}
G.~L. Kane, J.~Pumplin, and W.~Repko,
\newblock Phys. Rev. Lett. {\bf 41}, 1689 (1978).

\bibitem{Qiu:1991pp}
J.-w. Qiu and G.~F. Sterman,
\newblock Phys. Rev. Lett. {\bf 67}, 2264 (1991).

\bibitem{Kouvaris:2006zy}
C.~Kouvaris, J.-W. Qiu, W.~Vogelsang, and F.~Yuan,
\newblock Phys. Rev. D {\bf 74}, 114013 (2006), arXiv:hep-ph/0609238.

\bibitem{Kang:2010zzb}
Z.-B. Kang, F.~Yuan, and J.~Zhou,
\newblock Phys. Lett. B {\bf 691}, 243 (2010), arXiv:1002.0399.

\bibitem{Kang:2011hk}
Z.-B. Kang, J.-W. Qiu, W.~Vogelsang, and F.~Yuan,
\newblock Phys. Rev. D {\bf 83}, 094001 (2011), arXiv:1103.1591.

\bibitem{Metz:2012ct}
A.~Metz and D.~Pitonyak,
\newblock Phys. Lett. B {\bf 723}, 365 (2013), arXiv:1212.5037,
\newblock [Erratum: Phys.Lett.B 762, 549--549 (2016)].

\bibitem{Gamberg:2013kla}
L.~Gamberg, Z.-B. Kang, and A.~Prokudin,
\newblock Phys. Rev. Lett. {\bf 110}, 232301 (2013), arXiv:1302.3218.

\bibitem{Kanazawa:2014dca}
K.~Kanazawa, Y.~Koike, A.~Metz, and D.~Pitonyak,
\newblock Phys. Rev. D {\bf 89}, 111501 (2014), arXiv:1404.1033.

\bibitem{Gamberg:2017gle}
L.~Gamberg, Z.-B. Kang, D.~Pitonyak, and A.~Prokudin,
\newblock Phys. Lett. B {\bf 770}, 242 (2017), arXiv:1701.09170.

\bibitem{Qiu:1991wg}
J.-w. Qiu and G.~F. Sterman,
\newblock Nucl. Phys. B {\bf 378}, 52 (1992).

\bibitem{Kanazawa:2010au}
K.~Kanazawa and Y.~Koike,
\newblock Phys. Rev. D {\bf 82}, 034009 (2010), arXiv:1005.1468.

\bibitem{Cammarota:2020qcw}
Jefferson Lab Angular Momentum, J.~Cammarota {\em et~al.},
\newblock Phys. Rev. D {\bf 102}, 054002 (2020), arXiv:2002.08384.

\bibitem{Airapetian:2009ae}
HERMES, A.~Airapetian {\em et~al.},
\newblock Phys. Rev. Lett. {\bf 103}, 152002 (2009), arXiv:0906.3918.

\bibitem{Airapetian:2020zzo}
HERMES, A.~Airapetian {\em et~al.},
\newblock (2020), arXiv:2007.07755.

\bibitem{Adolph:2012sp}
COMPASS, C.~Adolph {\em et~al.},
\newblock Phys. Lett. {\bf B717}, 383 (2012), arXiv:1205.5122.

\bibitem{Adolph:2016dvl}
COMPASS, C.~Adolph {\em et~al.},
\newblock Phys. Lett. {\bf B770}, 138 (2017), arXiv:1609.07374.

\bibitem{Qian:2011py}
Jefferson Lab Hall A, X.~Qian {\em et~al.},
\newblock Phys. Rev. Lett. {\bf 107}, 072003 (2011), arXiv:1106.0363.

\bibitem{Bacchetta:2006tn}
A.~Bacchetta {\em et~al.},
\newblock JHEP {\bf 02}, 093 (2007), arXiv:hep-ph/0611265.

\bibitem{Brodsky:2002rv}
S.~J. Brodsky, D.~S. Hwang, and I.~Schmidt,
\newblock Nucl. Phys. B {\bf 642}, 344 (2002), arXiv:hep-ph/0206259.

\bibitem{Collins:2002kn}
J.~C. Collins,
\newblock Phys. Lett. B {\bf 536}, 43 (2002), arXiv:hep-ph/0204004.

\bibitem{Boer:2003cm}
D.~Boer, P.~Mulders, and F.~Pijlman,
\newblock Nucl. Phys. B {\bf 667}, 201 (2003), arXiv:hep-ph/0303034.

\bibitem{Aghasyan:2017jop}
COMPASS, M.~Aghasyan {\em et~al.},
\newblock Phys. Rev. Lett. {\bf 119}, 112002 (2017), arXiv:1704.00488.

\bibitem{Adamczyk:2015gyk}
STAR, L.~Adamczyk {\em et~al.},
\newblock Phys. Rev. Lett. {\bf 116}, 132301 (2016), arXiv:1511.06003.

\bibitem{Kang:2009bp}
Z.-B. Kang and J.-W. Qiu,
\newblock Phys. Rev. Lett. {\bf 103}, 172001 (2009), arXiv:0903.3629.

\bibitem{Anselmino:2016uie}
M.~Anselmino, M.~Boglione, U.~D'Alesio, F.~Murgia, and A.~Prokudin,
\newblock JHEP {\bf 04}, 046 (2017), arXiv:1612.06413.

\bibitem{Aschenauer:2016our}
E.-C. Aschenauer {\em et~al.},
\newblock (2016), arXiv:1602.03922.

\bibitem{Bland:2013pkt}
AnDY, L.~Bland {\em et~al.},
\newblock Phys. Lett. B {\bf 750}, 660 (2015), arXiv:1304.1454.

\bibitem{Abelev:2007ii}
STAR, B.~Abelev {\em et~al.},
\newblock Phys. Rev. Lett. {\bf 99}, 142003 (2007), arXiv:0705.4629.

\bibitem{Boer:2003tx}
D.~Boer and W.~Vogelsang,
\newblock Phys. Rev. D {\bf 69}, 094025 (2004), arXiv:hep-ph/0312320.

\bibitem{Bomhof:2007su}
C.~Bomhof, P.~Mulders, W.~Vogelsang, and F.~Yuan,
\newblock Phys. Rev. D {\bf 75}, 074019 (2007), arXiv:hep-ph/0701277.

\bibitem{Vogelsang:2007jk}
W.~Vogelsang and F.~Yuan,
\newblock Phys. Rev. D {\bf 76}, 094013 (2007), arXiv:0708.4398.

\bibitem{Qiu:2007ar}
J.-W. Qiu, W.~Vogelsang, and F.~Yuan,
\newblock Phys. Lett. B {\bf 650}, 373 (2007), arXiv:0704.1153.

\bibitem{Echevarria:2014xaa}
M.~G. Echevarria, A.~Idilbi, Z.-B. Kang, and I.~Vitev,
\newblock Phys. Rev. {\bf D89}, 074013 (2014), arXiv:1401.5078.

\bibitem{Bacchetta:2020gko}
A.~Bacchetta, F.~Delcarro, C.~Pisano, and M.~Radici,
\newblock (2020), arXiv:2004.14278.

\bibitem{Collins:2007nk}
J.~Collins and J.-W. Qiu,
\newblock Phys. Rev. D {\bf 75}, 114014 (2007), arXiv:0705.2141.

\bibitem{Rogers:2010dm}
T.~C. Rogers and P.~J. Mulders,
\newblock Phys. Rev. D {\bf 81}, 094006 (2010), arXiv:1001.2977.

\bibitem{bnltalk}
H.~Liu,
\newblock {\em {Talk given at RIKEN BNL Workshop Jet Observables at the
  Electron-Ion Collider}}, 2020.

\bibitem{Adare:2016bug}
PHENIX, A.~Adare {\em et~al.},
\newblock Phys. Rev. D {\bf 95}, 072002 (2017), arXiv:1609.04769.

\bibitem{Aidala:2018bjf}
PHENIX, C.~Aidala {\em et~al.},
\newblock Phys. Rev. D {\bf 98}, 072004 (2018), arXiv:1805.02450.

\bibitem{Sun:2014gfa}
P.~Sun, C.~P. Yuan, and F.~Yuan,
\newblock Phys. Rev. Lett. {\bf 113}, 232001 (2014), arXiv:1405.1105.

\bibitem{Sun:2015doa}
P.~Sun, C.~P. Yuan, and F.~Yuan,
\newblock Phys. Rev. {\bf D92}, 094007 (2015), arXiv:1506.06170.

\bibitem{Buffing:2018ggv}
M.~G.~A. Buffing, Z.-B. Kang, K.~Lee, and X.~Liu,
\newblock (2018), arXiv:1812.07549.

\bibitem{Chien:2019gyf}
Y.-T. Chien, D.~Y. Shao, and B.~Wu,
\newblock JHEP {\bf 11}, 025 (2019), arXiv:1905.01335.

\bibitem{Chien:2020hzh}
Y.-T. Chien {\em et~al.},
\newblock (2020), arXiv:2005.12279.

\bibitem{Kang:2020fka}
Z.-B. Kang, X.~Liu, S.~Mantry, and D.~Y. Shao,
\newblock (2020), arXiv:2008.00655.

\bibitem{Cacciari:2011ma}
M.~Cacciari, G.~P. Salam, and G.~Soyez,
\newblock Eur. Phys. J. C {\bf 72}, 1896 (2012), arXiv:1111.6097.

\bibitem{Bauer:2000ew}
C.~W. Bauer, S.~Fleming, and M.~E. Luke,
\newblock Phys. Rev. D {\bf 63}, 014006 (2000), arXiv:hep-ph/0005275.

\bibitem{Bauer:2000yr}
C.~W. Bauer, S.~Fleming, D.~Pirjol, and I.~W. Stewart,
\newblock Phys. Rev. {\bf D63}, 114020 (2001), arXiv:hep-ph/0011336.

\bibitem{Bauer:2001ct}
C.~W. Bauer and I.~W. Stewart,
\newblock Phys. Lett. {\bf B516}, 134 (2001), arXiv:hep-ph/0107001.

\bibitem{Bauer:2001yt}
C.~W. Bauer, D.~Pirjol, and I.~W. Stewart,
\newblock Phys. Rev. {\bf D65}, 054022 (2002), arXiv:hep-ph/0109045.

\bibitem{Bauer:2002nz}
C.~W. Bauer, S.~Fleming, D.~Pirjol, I.~Z. Rothstein, and I.~W. Stewart,
\newblock Phys. Rev. {\bf D66}, 014017 (2002), arXiv:hep-ph/0202088.

\bibitem{Kidonakis:1998nf}
N.~Kidonakis, G.~Oderda, and G.~F. Sterman,
\newblock Nucl. Phys. B {\bf 531}, 365 (1998), arXiv:hep-ph/9803241.

\bibitem{Chiu:2011qc}
J.-y. Chiu, A.~Jain, D.~Neill, and I.~Z. Rothstein,
\newblock Phys. Rev. Lett. {\bf 108}, 151601 (2012), arXiv:1104.0881.

\bibitem{Chiu:2012ir}
J.-Y. Chiu, A.~Jain, D.~Neill, and I.~Z. Rothstein,
\newblock JHEP {\bf 05}, 084 (2012), arXiv:1202.0814.

\bibitem{Collins:2011zzd}
J.~Collins,
\newblock Camb. Monogr. Part. Phys. Nucl. Phys. Cosmol. {\bf 32}, 1 (2011).

\bibitem{Bacchetta:2004jz}
A.~Bacchetta, U.~D'Alesio, M.~Diehl, and C.~Miller,
\newblock Phys. Rev. D {\bf 70}, 117504 (2004), arXiv:hep-ph/0410050.

\bibitem{Kelley:2010fn}
R.~Kelley and M.~D. Schwartz,
\newblock Phys. Rev. D {\bf 83}, 045022 (2011), arXiv:1008.2759.

\bibitem{Echevarria:2015usa}
M.~G. Echevarria, I.~Scimemi, and A.~Vladimirov,
\newblock Phys. Rev. D {\bf 93}, 011502 (2016), arXiv:1509.06392,
\newblock [Erratum: Phys.Rev.D 94, 099904 (2016)].

\bibitem{Liu:2020jjv}
X.~Liu, F.~Ringer, W.~Vogelsang, and F.~Yuan,
\newblock (2020), arXiv:2008.03666.

\bibitem{Aybat:2011zv}
S.~M. Aybat and T.~C. Rogers,
\newblock Phys. Rev. {\bf D83}, 114042 (2011), arXiv:1101.5057.

\bibitem{Kang:2011mr}
Z.-B. Kang, B.-W. Xiao, and F.~Yuan,
\newblock Phys. Rev. Lett. {\bf 107}, 152002 (2011), arXiv:1106.0266.

\bibitem{Sun:2013hua}
P.~Sun and F.~Yuan,
\newblock Phys. Rev. D {\bf 88}, 114012 (2013), arXiv:1308.5003.

\bibitem{Dai:2014ala}
L.-Y. Dai, Z.-B. Kang, A.~Prokudin, and I.~Vitev,
\newblock Phys. Rev. D {\bf 92}, 114024 (2015), arXiv:1409.5851.

\bibitem{Scimemi:2019gge}
I.~Scimemi, A.~Tarasov, and A.~Vladimirov,
\newblock JHEP {\bf 05}, 125 (2019), arXiv:1901.04519.

\bibitem{Moos:2020wvd}
V.~Moos and A.~Vladimirov,
\newblock (2020), arXiv:2008.01744.

\bibitem{Bacchetta:2005rm}
A.~Bacchetta, C.~Bomhof, P.~Mulders, and F.~Pijlman,
\newblock Phys. Rev. D {\bf 72}, 034030 (2005), arXiv:hep-ph/0505268.

\bibitem{Bomhof:2006dp}
C.~Bomhof, P.~Mulders, and F.~Pijlman,
\newblock Eur. Phys. J. C {\bf 47}, 147 (2006), arXiv:hep-ph/0601171.

\bibitem{Qiu:2007ey}
J.-W. Qiu, W.~Vogelsang, and F.~Yuan,
\newblock Phys. Rev. D {\bf 76}, 074029 (2007), arXiv:0706.1196.

\bibitem{Catani:2011st}
S.~Catani, D.~de~Florian, and G.~Rodrigo,
\newblock JHEP {\bf 07}, 026 (2012), arXiv:1112.4405.

\bibitem{Forshaw:2012bi}
J.~R. Forshaw, M.~H. Seymour, and A.~Siodmok,
\newblock JHEP {\bf 11}, 066 (2012), arXiv:1206.6363.

\bibitem{Rothstein:2016bsq}
I.~Z. Rothstein and I.~W. Stewart,
\newblock JHEP {\bf 08}, 025 (2016), arXiv:1601.04695.

\bibitem{Dasgupta:2001sh}
M.~Dasgupta and G.~P. Salam,
\newblock Phys. Lett. {\bf B512}, 323 (2001), arXiv:hep-ph/0104277.

\bibitem{Becher:2015hka}
T.~Becher, M.~Neubert, L.~Rothen, and D.~Y. Shao,
\newblock Phys. Rev. Lett. {\bf 116}, 192001 (2016), arXiv:1508.06645.

\bibitem{Becher:2016mmh}
T.~Becher, M.~Neubert, L.~Rothen, and D.~Y. Shao,
\newblock JHEP {\bf 11}, 019 (2016), arXiv:1605.02737,
\newblock [Erratum: JHEP05,154(2017)].

\bibitem{Sterman:2004en}
G.~F. Sterman,
\newblock Acta Phys. Polon. {\bf B36}, 389 (2005), arXiv:hep-ph/0410014.

\bibitem{Becher:2017nof}
T.~Becher, R.~Rahn, and D.~Y. Shao,
\newblock JHEP {\bf 10}, 030 (2017), arXiv:1708.04516.

\bibitem{Kang:2020yqw}
Z.-B. Kang, D.~Y. Shao, and F.~Zhao,
\newblock (2020), arXiv:2007.14425.

\bibitem{Dasgupta:2002bw}
M.~Dasgupta and G.~P. Salam,
\newblock JHEP {\bf 03}, 017 (2002), arXiv:hep-ph/0203009.

\bibitem{Balsiger:2019tne}
M.~Balsiger, T.~Becher, and D.~Y. Shao,
\newblock JHEP {\bf 04}, 020 (2019), arXiv:1901.09038.

\bibitem{Neill:2018yet}
D.~Neill,
\newblock JHEP {\bf 02}, 114 (2019), arXiv:1808.04897.

\bibitem{Banfi:2002hw}
A.~Banfi, G.~Marchesini, and G.~Smye,
\newblock JHEP {\bf 08}, 006 (2002), arXiv:hep-ph/0206076.

\bibitem{Hatta:2017fwr}
Y.~Hatta, E.~Iancu, A.~H. Mueller, and D.~N. Triantafyllopoulos,
\newblock JHEP {\bf 02}, 075 (2018), arXiv:1710.06722.

\bibitem{Zhu:2012ts}
H.~X. Zhu, C.~S. Li, H.~T. Li, D.~Y. Shao, and L.~L. Yang,
\newblock Phys. Rev. Lett. {\bf 110}, 082001 (2013), arXiv:1208.5774.

\bibitem{Li:2013mia}
H.~T. Li, C.~S. Li, D.~Y. Shao, L.~L. Yang, and H.~X. Zhu,
\newblock Phys. Rev. {\bf D88}, 074004 (2013), arXiv:1307.2464.

\bibitem{Angeles-Martinez:2018mqh}
R.~Angeles-Martinez, M.~Czakon, and S.~Sapeta,
\newblock JHEP {\bf 10}, 201 (2018), arXiv:1809.01459.

\bibitem{Liu:2014oog}
Z.~L. Liu, C.~S. Li, J.~Wang, and Y.~Wang,
\newblock JHEP {\bf 04}, 005 (2015), arXiv:1412.1337.

\bibitem{Moult:2015aoa}
I.~Moult, I.~W. Stewart, F.~J. Tackmann, and W.~J. Waalewijn,
\newblock Phys. Rev. D {\bf 93}, 094003 (2016), arXiv:1508.02397.

\bibitem{Bomhof:2004aw}
C.~Bomhof, P.~Mulders, and F.~Pijlman,
\newblock Phys. Lett. B {\bf 596}, 277 (2004), arXiv:hep-ph/0406099.

\bibitem{Bomhof:2007zz}
C.~J. Bomhof,
\newblock {Azimuthal Spin Asymmetries in Hadronic Processes},
\newblock Other thesis, Vrije U., 2007.

\bibitem{Gamberg:2010tj}
L.~Gamberg and Z.-B. Kang,
\newblock Phys. Lett. B {\bf 696}, 109 (2011), arXiv:1009.1936.

\bibitem{DAlesio:2017rzj}
U.~D'Alesio, F.~Murgia, C.~Pisano, and P.~Taels,
\newblock Phys. Rev. D {\bf 96}, 036011 (2017), arXiv:1705.04169.

\bibitem{DAlesio:2018rnv}
U.~D'Alesio, C.~Flore, F.~Murgia, C.~Pisano, and P.~Taels,
\newblock Phys. Rev. D {\bf 99}, 036013 (2019), arXiv:1811.02970.

\bibitem{Becher:2009qa}
T.~Becher and M.~Neubert,
\newblock JHEP {\bf 06}, 081 (2009), arXiv:0903.1126,
\newblock [Erratum: JHEP 11, 024 (2013)].

\bibitem{Sterman:2002qn}
G.~F. Sterman and M.~E. Tejeda-Yeomans,
\newblock Phys. Lett. B {\bf 552}, 48 (2003), arXiv:hep-ph/0210130.

\bibitem{Hornig:2017pud}
A.~Hornig, D.~Kang, Y.~Makris, and T.~Mehen,
\newblock JHEP {\bf 12}, 043 (2017), arXiv:1708.08467.

\bibitem{Ellis:2010rwa}
S.~D. Ellis, C.~K. Vermilion, J.~R. Walsh, A.~Hornig, and C.~Lee,
\newblock JHEP {\bf 11}, 101 (2010), arXiv:1001.0014.

\bibitem{Liu:2012sz}
X.~Liu and F.~Petriello,
\newblock Phys. Rev. D {\bf 87}, 014018 (2013), arXiv:1210.1906.

\bibitem{lapack99}
E.~Anderson {\em et~al.},
\newblock {\em {LAPACK} Users' Guide}, Third ed. (Society for Industrial and
  Applied Mathematics, Philadelphia, PA, 1999).

\bibitem{Kang:2015msa}
Z.-B. Kang, A.~Prokudin, P.~Sun, and F.~Yuan,
\newblock Phys. Rev. {\bf D93}, 014009 (2016), arXiv:1505.05589.

\bibitem{Su:2014wpa}
P.~Sun, J.~Isaacson, C.~P. Yuan, and F.~Yuan,
\newblock Int. J. Mod. Phys. {\bf A33}, 1841006 (2018), arXiv:1406.3073.

\bibitem{Becher:2016omr}
T.~Becher, B.~D. Pecjak, and D.~Y. Shao,
\newblock JHEP {\bf 12}, 018 (2016), arXiv:1610.01608.

\bibitem{Balsiger:2018ezi}
M.~Balsiger, T.~Becher, and D.~Y. Shao,
\newblock JHEP {\bf 08}, 104 (2018), arXiv:1803.07045.

\bibitem{Echevarria:2020hpy}
M.~G. Echevarria, Z.-B. Kang, and J.~Terry,
\newblock (2020), arXiv:2009.10710.

\bibitem{Abramowicz:2015mha}
H1, ZEUS, H.~Abramowicz {\em et~al.},
\newblock Eur. Phys. J. C {\bf 75}, 580 (2015), arXiv:1506.06042.

\bibitem{Kang:2019ctl}
Z.-B. Kang, A.~Prokudin, N.~Sato, and J.~Terry,
\newblock Comput. Phys. Commun. {\bf 258}, 107611 (2021), arXiv:1906.05949.

\bibitem{Currie:2017eqf}
J.~Currie {\em et~al.},
\newblock Phys. Rev. Lett. {\bf 119}, 152001 (2017), arXiv:1705.10271.

\bibitem{Aschenauer:2015eha}
E.-C. Aschenauer {\em et~al.},
\newblock (2015), arXiv:1501.01220.

\bibitem{Krohn:2012fg}
D.~Krohn, M.~D. Schwartz, T.~Lin, and W.~J. Waalewijn,
\newblock Phys. Rev. Lett. {\bf 110}, 212001 (2013), arXiv:1209.2421.

\bibitem{Waalewijn:2012sv}
W.~J. Waalewijn,
\newblock Phys. Rev. D {\bf 86}, 094030 (2012), arXiv:1209.3019.

\bibitem{Sjostrand:2007gs}
T.~Sjostrand, S.~Mrenna, and P.~Z. Skands,
\newblock Comput. Phys. Commun. {\bf 178}, 852 (2008), arXiv:0710.3820.

\end{thebibliography}

\bibliographystyle{h-physrev5}

\end{document}